\pdfoutput=1
\documentclass[12pt]{article}
\usepackage{graphicx}
\usepackage{amssymb,amsfonts,amsmath}

\def\lsim{\mathrel{\rlap{\lower3pt\hbox{\hskip1pt$\sim$}}
     \raise1pt\hbox{$<$}}} 
\def\gsim{\mathrel{\rlap{\lower3pt\hbox{\hskip1pt$\sim$}}
     \raise1pt\hbox{$>$}}} 
\newcommand{\beq}{\begin{equation}}
\newcommand{\eeq}{\end{equation}}
\newcommand{\bea}{\begin{eqnarray}}
\newcommand{\eea}{\end{eqnarray}}

\begin{document}
\title{Radiation of an electric charge in the field of a magnetic monopole}
\author{Michael Lublinsky$^{a,b}$, Claudia Ratti$^{a,c}$, Edward Shuryak$^a$\\
$^a$ {\small Department of Physics \& Astronomy, State University of New York}\\
{\small Stony Brook NY 11794-3800, USA}\\
$^b$ {\small Physics Department, Ben-Gurion University, Beer Sheva 84105, Israel}\\
$^c$ {\small Department of Theoretical Physics, University of Wuppertal, }\\ 
{\small Wuppertal 42119, Germany}}
\maketitle
\begin{abstract}
We consider the radiation of photons from quarks scattering on color-magnetic
monopoles in the Quark-Gluon Plasma. We consider a temperature
regime $T\gsim2T_c$, where monopoles can be considered as
static, rare objects embedded into
matter consisting mostly of the usual ``electric" quasiparticles, 
quarks and gluons. The calculation is performed in the
classical, non-relativistic approximation and results are compared to 
photon emission from Coulomb scattering of quarks,
known to provide a significant contribution to the
photon emission rates from QGP. The present study is a first step 
towards understanding whether this scattering process can give a sizeable contribution
to dilepton production in heavy-ion collisions. Our results are
encouraging: by comparing the magnitudes of the photon emission rate 
for the two processes,  we find a dominance in the case of quark-monopole scattering.
Our results display strong sensitivity to  finite densities of quarks and monopoles.
\end{abstract}
\section{Introduction}
Creating and studying Quark-Gluon Plasma, the deconfined phase of QCD, in the
laboratory has been the goal of experiments at CERN SPS and 
at the Relativistic Heavy Ion Collider (RHIC) facility in Brookhaven
National Laboratory, soon to be continued by the ALICE (and, to a smaller extent, by the two other collaborations)
at the Large Hadron Collider (LHC). Dileptons and photons are a particularly interesting
observable from heavy ion collisions, since electromagnetic
probes do not interact with the medium after their production, thus
carrying information about $all$  stages of the evolution \cite{Shuryak:1978ij}.
Discussion of dilepton production in
heavy ion collision experiments at CERN SPS \cite{CERES,NA60} and at RHIC \cite{PHENIX} 
and their comparison to theory can be found in \cite{Dusling:2007su,Turbide:2007mi,Rapp:2009yu},
for a recent review
see also e.g. \cite{Lapidus:2009zz} and references therein. The main
contributions to the production rates considered so far are hadronic decays in the mixed and
hadronic phases of the collision (the so-called hadronic cocktail), and
quark-antiquark annihilation in the QGP phase. After a long history of experimental studies of dileptons produced by 
charm decay, NA50/60 experiments finally concluded \cite{Arnaldi:2008er} that they do see QGP radiation, at intermediate dilepton
mass range 1-3 GeV, as predicted in   \cite{Shuryak:1978ij,Rapp:1999zw}.

However, the experimentally observed excess in dilepton production at small $p_t$ and small invariant dilepton mass (below $m_\rho\,\simeq\,0.77$ GeV)
remains a  puzzle: 
 the sum of all known contributions fails to explain the data by a large margin. 
Motivated by this, we search for additional, unexplored
mechanisms which might contribute to the dilepton production rate.
In particular, we will focus here on the role played by
color-magnetic monopoles: we want to estimate the contribution
to dilepton production  from quarks
which scatter on them. This methodological paper is our initial step towards an exploration of the subject: by no means
we claim any resolution of the puzzle.

The so called ``magnetic scenario" for QGP  \cite{Liao:2006ry} suggested that  the near-$T_c$ region 
is dominated by
monopoles. More specifically, these authors suggested to look at the magnetic sector as a (magnetic)
Coulomb plasma of monopoles in its liquid form. A line of
lattice-based results has led  to a very
similar conclusion~\cite{Chernodub:2006gu}.
This scenario has met with initial success by providing
an  explanation of the low viscosity observed at RHIC \cite{Liao:2006ry,RS} due
 to the large transport cross section induced by scattering on monopoles. 

Lattice monopoles are defined by the procedure 
\cite{DeGrand:1980eq} which locates the ends
of singular Dirac strings by calculating the total magnetic flux through
the boundary of  elementary 3-d boxes.
Since this depends on a certain gauge fixing,
for decades sceptics kept the viewpoint that  
those objects are just unphysical  UV gauge  noise.
Yet, many specific observables  -- e.g. monopole density   -- produced very
reasonable and consistent results,  apparently 
independent of the particular lattice parameters \cite{Chernodub:2006gu,D'Alessandro:2007su}.   
More recent results on monopole correlations  \cite{D'Alessandro:2007su}
quantitatively support the Coulomb plasma picture
of Ref.~\cite{Liao:2006ry},  providing further reasons to think that
 monopoles are not artefacts but meaningful physical objects,
present in the QGP as a  source of a Coulomb-like magnetic field on which charged particles (quarks) can scatter.
We work under the same assumption in the present paper.
 We are not advocating this magnetic scenario, but rather use it to
 estimate the contribution of radiation on monopoles.
 We will not need any assumption about monopole coupling, internal structure or correlations, only their density.

 Although we were motivated by 
the dilepton puzzle and will eventually 
aim at solving it as a final goal, we start from the simpler problem of soft photon radiation during the collision process.
The emission of real photons is a process which is closely related to dilepton production: the latter takes place through the emission of a virtual photon.
 In the QGP phase, the leading perturbative diagram is the
  Compton-like process
 ($qg\rightarrow q\gamma$ and the crossing diagram $q\bar q\rightarrow \gamma g$), 
while perturbatively subleading  bremsstrahlung diagrams 
($qq\rightarrow qq\gamma$)
and LPM-type resummed effects are in fact  equally important \cite{Aurenche:1998nw}. 

The classical trajectory of a particle with electric charge\footnote{Here and below we call $e$ the strong interaction coupling constant, using
QED-like field normalization of the fields and $4\pi$ for consistency with the textbook material we use: note that
$e^2=\alpha_s$. The name $g$ is reserved to color-magnetic coupling, while
the electromagnetic coupling
will be denoted as $e_{em}$, again with $e_{em}^2=\alpha_{em}$. } $e$
in the field of an infinitely heavy monopole with magnetic charge $g$ takes place on the surface of a cone. 
The static monopole approximation is valid for a regime of temperatures $T\gsim2T_c$, where they
can be considered heavy, rare objects embedded into matter consisting mostly of the usual 
electric quasiparticles, quarks and gluons.
The Lorentz force 
acting on the electric charge is proportional to the product of both couplings $(eg)$. 
Thanks to the Dirac charge quantization condition, $eg=1$ and thus it is not 
a small parameter and it is $T$-independent.

As a first step towards the solution of the problem, in the present paper we compute the radiation from  a non-relativistic electrically charged particle moving in the field of a monopole along the $classical$ trajectory, ignoring back reaction.  A full quantum and relativistic study is postponed for future investigation.  
Below we will discuss the applicability limits of our approximation. Therefore,  our present calculation  cannot  address the actual phenomenological questions yet,  but we 
can get an insight
into how sizeable the effect of monopoles can be.  To this purpose, we will compare our results with the parallel computation of photon emission rate in the process of Coulomb scattering of quarks, in the same approximations, 
regarding the Coulomb problem as a benchmark for comparison.   

 Let us start with a  ``naive estimate" for the ratio of emission rates
of quark-monopole vs Coulomb scattering of quarks:
\beq
{I^{qM}\over I^{qq}}\sim\, { (eg)^2 v^2\over e^4}\, {\mu\over m} \,{n_M\over n_q }.
\eeq
First, the emission amplitude from  monopoles is suppressed  by the velocity $v$ of the incoming electric particle, because the underlying
scattering happens due to Lorentz rather than Coulomb force. Second, it obviously contains 
the density of monopoles $n_M\sim T^3/\ln^3(T)$, which is 
smaller than the quark density $n_q\sim T^3$.

On the other hand, this rate is enhanced by the ratio of coupling constants.
The numerator includes the product
of the electric gauge coupling constant $e$ and the magnetic one $g$:
in the units we are using, the Dirac quantization condition implies $eg=1$ (actually $\hbar$), while
the electric Coulomb scattering is proportional to  $e^2=\alpha_s\sim 1/\ln(T) \ll 1$.
The  $\ln^2(T)$ in the numerator to a significant extent compensates the smallness of the monopole density $n_M\sim 1/\ln^3(T)$. 
Although formally still decreasing at large $T$, as is the corresponding ratio of the contributions to viscosity \cite{RS}, large angle and even
backward scattering induced by monopoles may make this ratio numerically enhanced. 
 The reduced mass $\mu=m/2$  enters the Coulomb scattering problem whereas, in the limit of infinite monopole mass, the quark mass $m$ appears in Newton law.
There is an additional relative
enhancement in favour of monopoles, which is due to the
Casimirs of the $qq$ and $q\bar q$ potentials. We will work out these factors in the following;
they roughly bring a factor
1/4 compared to the $q-M$ scattering.  For typical (thermal) velocity
of 0.7, $\alpha_s=0.8$, and the ratio of densities 0.2 we end up
with a relative effect of order 1/2. This is obviously a good start, suggesting it is worth to examine the problem in more details.  

We will see below that the above crude estimate is indeed correct when the velocity of quarks is sufficiently close to one (we will still be using the non-relativistic approximation); it  is also correct in the ultrarelativistic  limit.  There is, however,  a very important effect, which noticeably enhances the relative contribution of monopoles.  The above estimate holds for a fixed impact parameter.  For soft radiation, only emission from large impact parameters is relevant (and this is the only region where our approximations are in fact valid).  However, finite densities
of quarks and monopoles in the plasma provide  natural cutoffs for maximal impact parameters. Moreover,  since the density of monopoles in our temperature
regime is much smaller 
than the density  of quarks, the relevant upper cutoff on the impact parameter for the Coulomb problem is much smaller  and it leads to a significant suppression of  the Coulomb-induced emission rate  relatively to the rate due to scattering on monopoles.  

Our final results are very encouraging. We find that the soft photon emission rate from quark-monopole scattering could  be as large as the mechanisms previously accounted for. Therefore, the problem calls for further and more detailed investigation at the full quantum level, which will be our next step in this project.

The paper is organized as follows. In Section 2 we provide a brief overview of the
classical motion of an electric charge in a Coulomb magnetic field: even if these results
are well-known, a short summary is useful, since we will make large use of them in the following. 
In Section 3 we compute the photon radiation rate for an electric particle moving in the field of a static monopole. This section can be regarded as an extra chapter for Ref. \cite{LL2},
in which radiation rates for several trajectories were computed. In Section 4 we present
an estimate of our effect in the case of quarks scattering on monopoles in the QGP:
we use parameter estimates which are typical of the QGP produced at RHIC.
We compare our results with those obtained in the Coulomb problem, and discuss
the validity of our approximations.
We draw our conclusions in the last Section (5), where we also indicate   future improvements. Appendix A supplements Section 3 by providing details of analytical
computations  for the emission rate. Appendix B presents a calculation of the quark density in the PNJL model, a result which is needed when we apply our calculations to QGP.

\section{Classical quark-monopole scattering}
\begin{center}
\begin{figure}[b]
\hspace{3cm}
\parbox{6cm}{
\scalebox{.81}{
\includegraphics{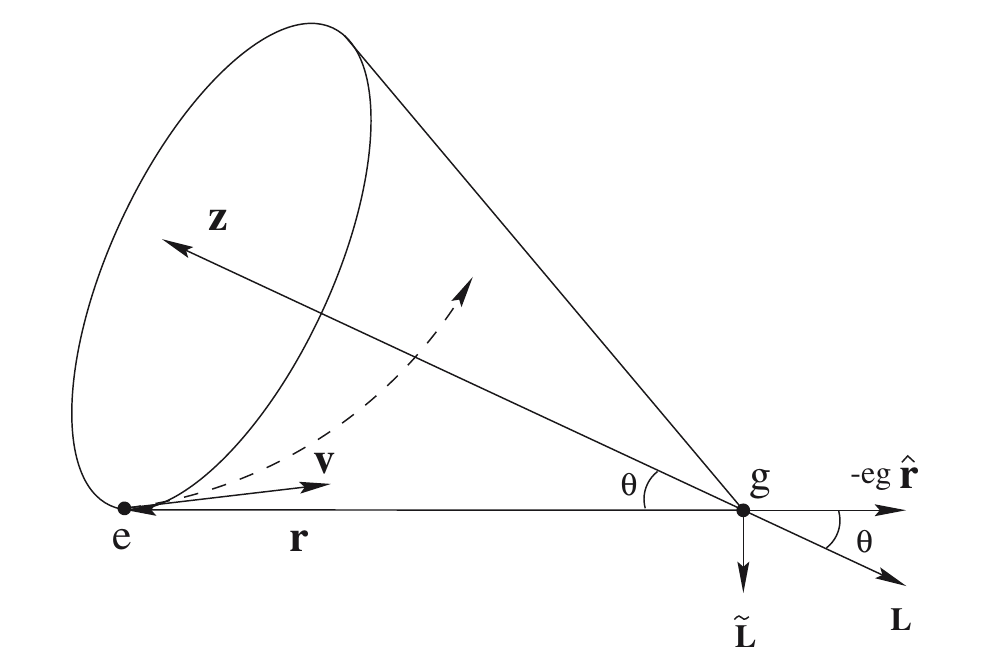}}}
\caption{The motion of an electric charge in the field of a magnetic monopole.}
\label{fig1}
\end{figure}
\end{center}

\begin{figure}
\hspace{-.8cm}
\begin{minipage}{.33\textwidth}
\vspace{1cm}
\parbox{6cm}{
\scalebox{.45}{
\includegraphics{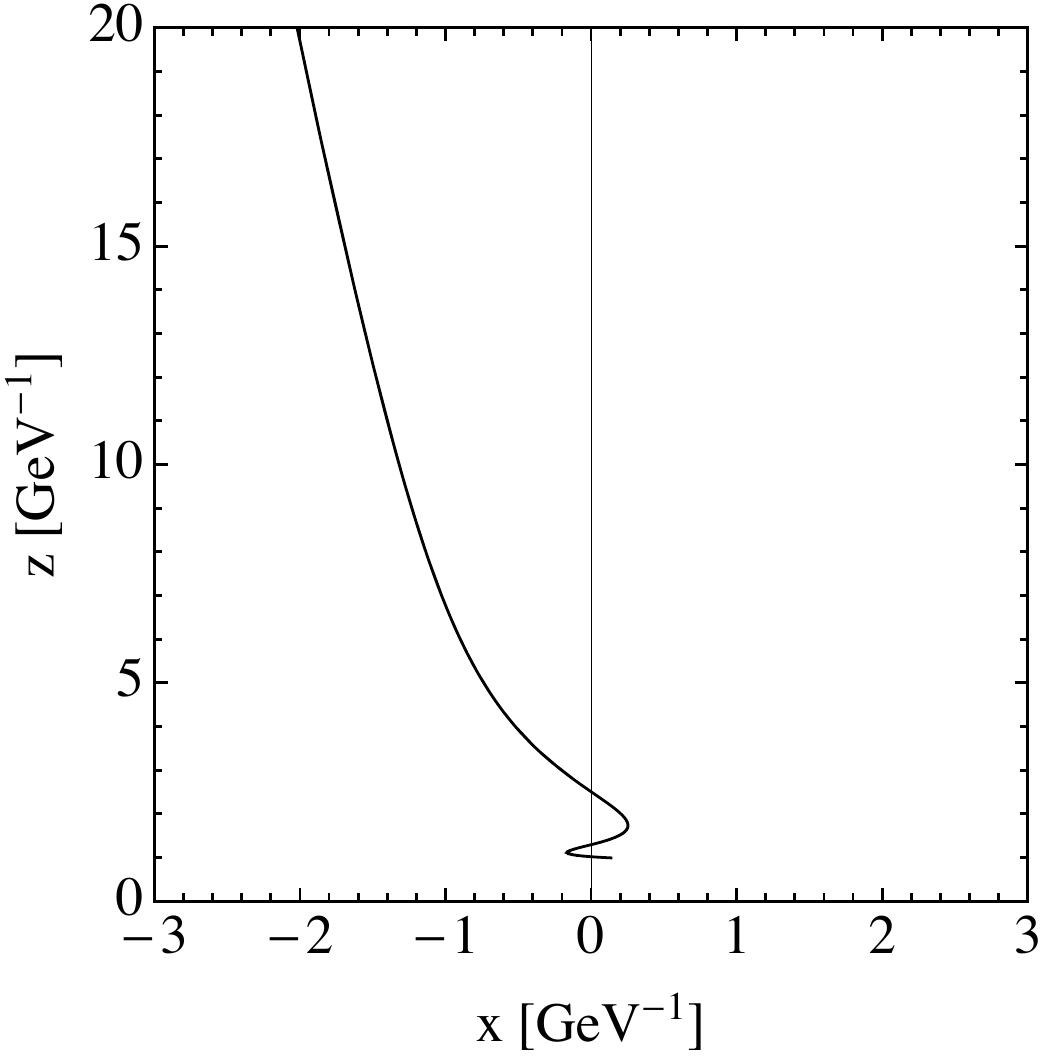}\\}}
\centerline{(a)}
\end{minipage}
\hspace{-.4cm}
\begin{minipage}{.33\textwidth}
\vspace{1cm}
\parbox{6cm}{
\scalebox{.45}{
\includegraphics{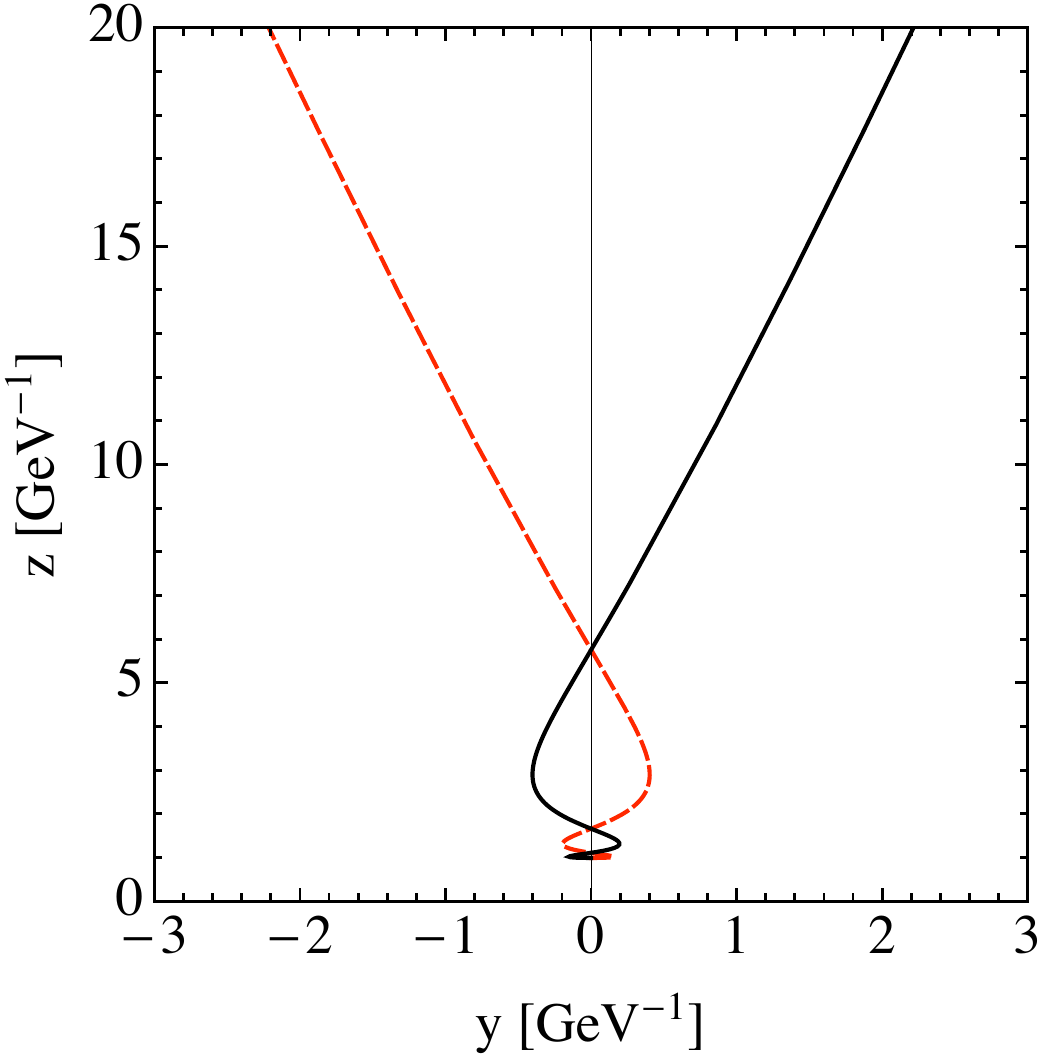}\\}}
\centerline{(b)}
\end{minipage}
\hspace{-.3cm}
\begin{minipage}{.33\textwidth}
\parbox{6cm}{
\scalebox{.6}{
\includegraphics{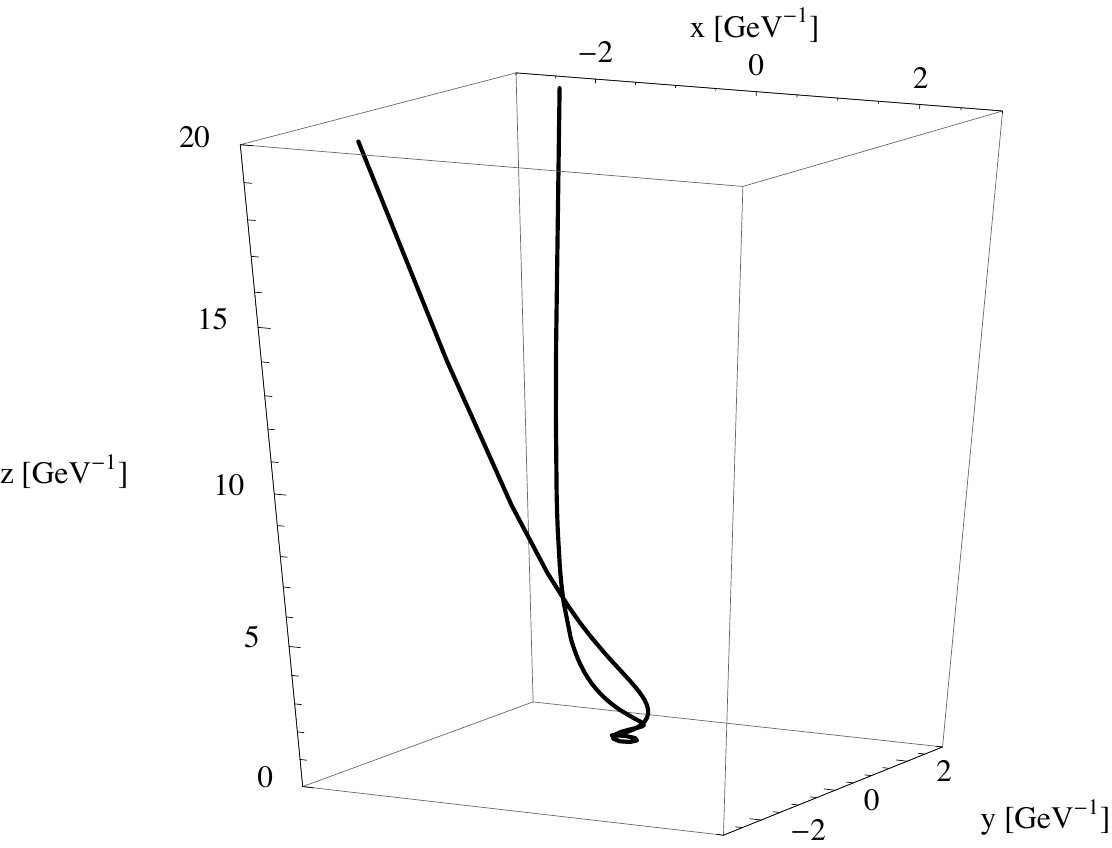}\\}}
\centerline{(c)}
\end{minipage}
\caption{Example of electric charge trajectory in the field of a magnetic monopole. (a) $x$ and $z$ axes: the trajectories of the incoming particle (from $t=-\infty$ to $t=0$) and the outgoing particle (from $t=0$ to $t=+\infty$) overlap in this case. (b) $y$ and $z$ axes. Solid line: incoming particle (from $t=-\infty$ to $t=0$). Dashed: outgoing particle (from $t=0$ to $t=+\infty$).
(c) Three dimensional trajectory. For all plots we use $v=0.5;~b=1$ GeV$^{-1}$, $m=0.3$ GeV.}
\label{trajectory}
\end{figure}

We consider the classical, non-relativistic motion of a charge in an external 
field \cite{Shnir:2005xx,Milton:2006cp,Boulware:1976tv}.
A pointlike magnetic charge $g$ is the source of a Coulomb-like magnetic field
\beq
\vec{B}=g\frac{\vec{r}}{r^3}.
\eeq
The equation of motion of an electrically charged particle $e$ in such a field
is
\beq
m\frac{d^2\vec{r}}{dt^2}=e\vec{v}\times\vec{B}=\frac{eg}{r^3}\frac{d\vec{r}}{dt}\times\vec{r};
\label{lorentz}
\eeq
the static monopole is located at the origin and the vector $\vec{r}$ defines the position of the
electric charge (see Fig. \ref{fig1}). In the following, we set $c=1$ for simplicity. 
We also use the convention $e^2=\alpha$, and therefore $eg=1$.
In this process, the kinetic energy of the electric charge is a constant:
\beq
E=\frac{mv^2}{2}=\mathrm{const.},
\eeq
as is the absolute value of the velocity vector $v$. There is no closed orbit in the charge-monopole
system: the electric charge is falling down from infinitely far away onto the monopole,
approaching a minimal distance $b$ and being reflected back to infinity. This is evident from the
trajectory
\beq
r=\sqrt{v^2t^2+b^2}.
\label{r}
\eeq
A special feature of such a motion is that the conserved angular momentum is different from the
ordinary case: the absolute value of the ordinary angular momentum is conserved, but its direction
is not constant. The generalized angular momentum
\beq
\vec{J}=\left[\vec{r}\times m\vec{v}\right]-eg\frac{\vec{r}}{r}
\eeq
is an integral of motion.
The trajectory of the electric charge does not lie in the plane of scattering orthogonal to the
angular momentum; the angle between the vectors $\vec{J}$ and $\vec{r}$ is a constant and
the electric charge is moving on the surface of a cone whose axis is directed along $-\vec{J}$
with the cone angle $\theta$ defined as:
\beq
\sin\theta=\frac{mvb}{\sqrt{\left(mvb\right)^2+\left(eg\right)^2}},~~~~~~~~~
\cos\theta=\frac{eg}{\sqrt{\left(mvb\right)^2+\left(eg\right)^2}}.
\eeq
The velocity of the electrically charged particle is
\beq
\vec{v}=\frac{d\vec{r}}{dt}=\frac{1}{mr^2}\left[\vec{J}\times\vec{r}\right]
+\frac{v^2t}{r}\hat{r}=\frac{1}{mr^2}\left[\vec{J}\times\vec{r}\right]
+\frac{v}{\sqrt{1+\left(b/vt\right)^2}}\hat{r}=\vec{v}_\varphi\times\vec{r}+v_r\hat{r}
\label{velocity}
\eeq
where the angular and radial components of the velocity vector are
\beq
\vec{v}_\varphi=\frac{\vec{J}}{mr^2},~~~~~~~~~~~~v_r=\frac{v}{\sqrt{1+\left(b/vt\right)^2}};
\eeq
asymptotically we have
\beq
v_\varphi\Big |_{t=\pm\infty}=0,~~~~~~~~~~~~v_r\Big |_{t=\pm\infty}=v,
\eeq
while at the turning point of the path we have
\beq
v_\varphi\Big |_{t=0}=\frac{\sqrt{\left(mvb\right)^2+\left(eg\right)^2}}{mb^2},~~~~~~~~~~~~
v_r\Big |_{t=0}=0.
\eeq
The azimuthal angle $\varphi$ as a function of time can be obtained by integrating the angular velocity:
\beq
\varphi=\frac{1}{\sin\theta}\arctan\frac{vt}{b}.
\eeq
An example of trajectory is shown in Fig.~\ref{trajectory}.

The acceleration of the electric charge, $\vec{a}=\frac{d^2\vec{r}}{dt^2}$, follows from Eqs.~(\ref{lorentz}) and (\ref{velocity}):
\beq
\vec{a}=\frac{eg}{mr^3}\vec{v}\times\vec{r}=\frac{eg}{mr^3}\left(\vec{v}_\varphi\times\vec{r}\right)
\times\vec{r}.
\eeq
We recall that $\vec{v}_\varphi$ is directed along the $z$ axis (the cone axis) so that:
\bea
\vec{a}&=&\frac{eg}{m}\frac{(v_\varphi)_z}{r^3}\left[r_xr_z,r_yr_z,-(r_{x}^2+r_{y}^{2})\right]
\nonumber\\
&=&\frac{eg}{m^2r^3}\sqrt{(mvb)^2+(eg)^2}\sin\theta\cos\theta\left[-\cos\varphi,-\sin\varphi,-\tan\theta\right].
\label{acc}
\eea
\section{Radiation}
Following Ref. \cite{LL2} we define
the intensity $dI$ of radiation into the element of solid angle $d\Omega$ as the
amount of energy passing in unit time through the element $df=R_{0}^{2}d\Omega$ of
the spherical surface with center at the origin and radius $R_0$. This quantity is equal to the energy flux density ($\vec{S}=\frac{H^2}{4\pi}\vec{n}$) multiplied by $df$:
\beq
dI=\frac{H^2}{4\pi}R_{0}^{2}d\Omega.
\label{intensity1}
\eeq
The energy $d\mathcal{E}_{\vec{n}\omega}$, radiated into the element of solid angle $d\Omega$
in the form of waves with frequencies in the interval $d\omega/2\pi$ is obtained from Eq.
(\ref{intensity1}) by replacing the square of the field by the square modulus of
its Fourier components and multiplying by 2:
\beq
d\mathcal{E}_{\vec{n}\omega}=\frac{|\vec{H}_{\omega}|^2}{2\pi}R_{0}^{2}d\Omega
\frac{d\omega}{2\pi}
\label{e}
\eeq
where
\beq
\vec{H}_{\omega}=i\vec{k}\times\vec{A}_{\omega}
\eeq
and
\beq
\vec{A}_{\omega}=\frac{e^{ikR_0}}{R_0}\int_{-\infty}^{+\infty}e_{em}\vec{v}(t)e^{i(\omega t-\vec{k}
\cdot\vec{r}(t))}dt.
\eeq

We work in the dipole approximation, namely we neglect retardation effects. This
approximation is valid provided that $v\ll c$. In this case, the field can be
considered as a plane wave, and therefore in determining the field it is
sufficient to calculate the vector potential. 
The intensity of the dipole radiation is:
\beq
dI=\frac{\left(e_{em}\vec{a}\right)^2}{4\pi}\sin^2\theta d\Omega,
\eeq
where $\vec{a}$ is the acceleration of the electric charge. Replacing $d\Omega=2\pi\sin\theta d\theta$ and integrating over $\theta$ from $0$ to
$\pi$, we find the total radiation:
\beq
I=\frac23(e_{em}\vec{a})^2.
\label{intensity}
\eeq
The quantity $d\mathcal{E}_{\omega}$ of energy radiated throughout the time of the
collision in the form of waves with frequencies in the interval $d\omega/2\pi$ is obtained from
Eq. (\ref{intensity}) by replacing $\vec{a}$ by its Fourier component $\vec{a}_\omega$ and
multiplying by 2:
\beq
d\mathcal{E}_{\omega}=\frac{4}{3}\left(e_{em}\vec{a}_{\omega}\right)^2\frac{d\omega}{2\pi}.
\eeq

In a standard planar collision, the total radiation $d\kappa_\omega$ in a given frequency interval $d\omega$ can be obtained
by multiplying the radiation $d\mathcal{E}_{\omega}$ from a single particle (with given impact parameter $\rho$) by the measure $2\pi \rho d\rho$ and integrating over $\rho$ from $\rho_{min}$ to
$\rho_{max}$. 
In our case, the motion of the electric particle takes place on the surface of a cone,
therefore we have to find the corresponding measure for our process.

If we project the conical
motion of the electric particle on a plane orthogonal to $\vec{J}$, it is possible to define
the standard impact parameter $\rho$ for the planar motion \cite{Boulware:1976tv}.
We can define a position vector $\vec{R}$:
\beq
\vec{R}=\frac{\hat{J}\times(\vec{r}\times\hat{J})}{\cos\theta}=\frac{1}{\cos\theta}\left[\vec{r}
-\hat{J}\left(\vec{r}\cdot\hat{J}\right)\right],
\eeq
which is the projection of $\vec{r}$ onto the plane perpendicular to $\vec{J}$, times a factor
$1/\cos\theta$ chosen so that $\vec{R}$ and $\vec{r}$ have the same length.
At $t=-\infty$ we have:
\beq
\dot{\vec{R}}_{-\infty}=\frac{1}{\cos\theta}\left[\vec{v}
-\hat{J}\left(\vec{v}\cdot\hat{J}\right)\right].
\eeq
The mechanical angular momentum of the projected motion is the same as the
conserved total angular momentum of the motion in the monopole field:
\beq
\vec{J}=m\vec{R}\times\dot{\vec{R}};~~~~~~~~~~~~~
|\vec{J}|=m\rho|\dot{\vec{R}}_{-\infty}|.
\eeq
From the above Equation we get:
\beq
\rho=\frac{\sqrt{\left(mvb\right)^2+\left(eg\right)^2}}{mv}
\eeq
from which it is easy to obtain that $\rho d\rho=bdb$:
the measure over which we need to integrate $d\mathcal{E}_\omega$ turns out to be
equal to $bdb$ for the conical motion. Therefore, we can identify $b$ as the real impact parameter
for the process that we are considering ($b$ is actually the minimal distance between the
electric particle and the monopole, which is reached at $t=0$):
\beq
\frac{d\kappa_\omega}{d\omega}=\int_{b_{min}}^{b_{max}}2\pi bdb{d\mathcal{E}_\omega\over d\omega}.
\eeq
The limits of integration for $b$ are related to the specific scattering process we are dealing with. 
For example, if we consider a single quark scattering on a single monopole, we have
$b_{max}\rightarrow\infty$, while $b_{min}$ can be identified with the size of the monopole core.
In the problem we are considering, in which there is a finite density of monopoles, $b_{max}$ turns out to be finite, as we will see in the next Section.

We start by calculating the total radiation throughout the time of the collision following 
Eq. (\ref{intensity}):
\beq
\mathcal{I}=\int_{-\infty}^{\infty}I(t)dt=\frac23\alpha_{em}\int_{-\infty}^{\infty}|\vec{a}(t)|^2dt;
\eeq
the acceleration components in the coordinate space are given in Eq. (\ref{acc}).
We obtain:
\beq
I(t)=\frac23\alpha_{em}\frac{(eg)^2v^2b^2}{m^2r(t)^6}=\frac23\alpha_{em}\frac{(eg)^2v^2b^2}{m^2(v^2t^2+b^2)^3}
\eeq
so that:
\beq
\mathcal{I}=\frac23\alpha_{em}\frac{(eg)^2v}{m^2b^3}\int_{-\infty}^{\infty}\frac{d\tau}{(\tau^2+1)^3}=
\frac\pi4\alpha_{em}\frac{(eg)^2v}{m^2b^3};
\eeq
the total radiation can be obtained in the following way:
\beq
\kappa=\int_{b_{min}}^{b_{max}}2\pi bdb\mathcal{I}=\frac{\pi^2}{2}\alpha_{em}\frac{(eg)^2v}{m^2}
\left(\frac{1}{b_{min}}-\frac{1}{b_{max}}\right).
\eeq
We now proceed by calculating the Fourier transform of the acceleration $\vec{a}$
in the case of quark-monopole scattering:
\bea
(a_{x})_{\omega}&=&-\frac{\left(eg\right)^2}{m^2b^2v\xi}
\int_{-\infty}^{\infty}dt\frac{\exp\left[i \bar{\omega} t\right]\cos\left[\xi\mathrm{arctan} t\right]}{\left(t^2+1\right)^{3/2}}
\\
(a_{y})_{\omega}&=&-\frac{\left(eg\right)^2}{m^2b^2v\xi}
\int_{-\infty}^{\infty}dt\frac{\exp\left[i \bar{\omega} t\right]\sin\left[\xi\mathrm{arctan} t\right]}{\left(t^2+1\right)^{3/2}}
\\
(a_{z})_{\omega}&=&-\frac{\left(eg\right)^2}{mb\xi}
\int_{-\infty}^{\infty}dt\frac{\exp\left[i \bar{\omega} t\right]}{\left(t^2+1\right)^{3/2}}
\eea
where
\beq
\xi=\frac{\sqrt{(mvb)^2+(eg)^2}}{mvb},~~~~~~~~~~~~\bar{\omega}=\omega\frac{b}{v}.
\eeq
For positive $\omega$ we obtain (see Appendix A):
\bea
(a_{x})_{\omega}&=&\frac{(eg)^2}{m^2b^2v\xi}\left\{\exp\left(-\bar{\omega}\right)\cos\left(\frac{\pi\xi}{2}\right)
\left[\frac14\Gamma\left(\frac12\left(\xi-1\right)\right)U\left(\frac12\left(\xi-1\right),-1,2\bar{\omega}\right)\right.\right.
\nonumber\\
\nonumber\\
&\!\!\!\!\!\!\!\!\!\!\!\!\!\!\!\!\!\!\!\!\!\!\!\!\!\!\!\!\!\!\!\!\!\!\!\!\!\!\!\!\!\!\!\!\!\!+&\!\!\!\!\!\!\!\!\!\!\!\!\!\!\!\!\!\!\!\!\!\!\!\!\!\!\!\!\left.\left.\frac{4p!\Gamma\left(-p+\frac{\xi+3}{2}\right)}{\xi^2-1}\sum_{k=0}^{p}\frac{\left(-\bar{\omega}\right)^k}{k!\left(p-k\right)!\Gamma\left(\frac{\xi-3}{2}-p+k+1\right)}\right.\right.
\times
\label{wx}\\
\nonumber\\
&&~~~~~~~~~~~~~~~~~~~~~~\times\left.\left.2^{k-2}
\Gamma\left(p-\frac{\xi+1}{2}\right)U\left(p-\frac{\xi+1}{2},k-1,2\bar{\omega}\right)
\right]\right\}
\nonumber\\
\nonumber\\
\nonumber\\
(a_{y})_{\omega}&=&-\frac{(eg)^2}{m^2b^2v\xi}\left\{i\exp\left(-\bar{\omega}\right)\cos\left(\frac{\pi\xi}{2}\right)
\left[\frac14\Gamma\left(\frac12\left(\xi-1\right)\right)U\left(\frac12\left(\xi-1\right),-1,2\bar{\omega}\right)\right.\right.
\nonumber\\
\nonumber\\
&\!\!\!\!\!\!\!\!\!\!\!\!\!\!\!\!\!\!\!\!\!\!\!\!\!\!\!\!\!\!\!\!\!\!\!\!\!\!\!\!\!\!\!\!\!-&\!\!\!\!\!\!\!\!\!\!\!\!\!\!\!\!\!\!\!\!\!\!\!\!\!\!\!\!\left.\left.\frac{4p!\Gamma\left(-p+\frac{\xi+3}{2}\right)}{\xi^2-1}\sum_{k=0}^{p}\frac{\left(-\bar{\omega}\right)^k}{k!\left(p-k\right)!\Gamma\left(\frac{\xi-3}{2}-p+k+1\right)}\right.\right.
\times
\label{wy}\\
\nonumber\\
&&~~~~~~~~~~~~~~~~~~~~~~~\times\left.\left.2^{k-2}
\Gamma\left(p-\frac{\xi+1}{2}\right)U\left(p-\frac{\xi+1}{2},k-1,2\bar{\omega}\right)
\right]\right\}
\nonumber\\
\nonumber\\
\nonumber\\
\left(a_z\right)_\omega&=&-\frac{2eg\bar{\omega}}{mb\xi}K_1\left(\bar{\omega}\right),
\label{wz}
\eea
where $p$ is the smallest integer number larger than $(\xi+3)/2$. $p-2$ is the number of full rotations
around $z$ axis. 
$U(a,b,z)$ is the confluent hypergeometric function with integral representation:
\beq
U(a,b,z)=\frac{1}{\Gamma(a)}\int_{0}^{\infty}e^{-zt}t^{a-1}(1+t)^{b-a-1}dt
\eeq
and $K_n(z)$ is the modified Bessel function of the second kind:
\beq
K_n(z)=\frac{\Gamma(n+\frac12)(2z)^n}{\sqrt{\pi}}\int_{0}^{\infty}\frac{\cos t~dt}{\left(t^2+z^2\right)^{n+1/2}}.
\eeq
Equations (\ref{wx})-(\ref{wy}) are strictly valid for $\omega\ge0$ and for any value of $\xi$, except odd, integer
numbers. When $\xi$ is an odd integer number, the above formulas vanish identically, and
the integral is given by Eq. ({\ref{oddxi}) (see Appendix A).
The behavior of $(a_x)_\omega$, $(a_y)_\omega$ and $(a_z)_\omega$ as functions of
$\omega$ and $b$ is shown in the two panels of Fig.~\ref{a}.
The component $(a_y)_\omega$ is purely imaginary and vanishes at $\omega=0$.
The values of the acceleration components at $\omega=0$ are:
\bea
(a_x)_{\omega=0}=\frac{2v}{\xi}\cos\left[\frac{\pi\xi}{2}\right]\,;\ \ \ \ \ \ \ \ \ \ \ \ (a_y)_{\omega=0}=0\,;
\ \ \ \ \ \ \ \ \ \ \ \ \ 
(a_z)_{\omega=0}=-\frac{(eg)}{mb\xi}.
\eea
The subleading small $\omega$ asymptotic behavior can be found in Appendix A.
The photon emission rate is  finite as $\omega\rightarrow 0$.
It is of course how it should be: the corresponding number of photons $dN_{\omega}=dI_{\omega}/\hbar \omega$ 
would show standard logarithmic IR divergence.

\begin{figure}
\hspace{-.8cm}
\begin{minipage}{.48\textwidth}
\parbox{6cm}{
\scalebox{.73}{
\includegraphics{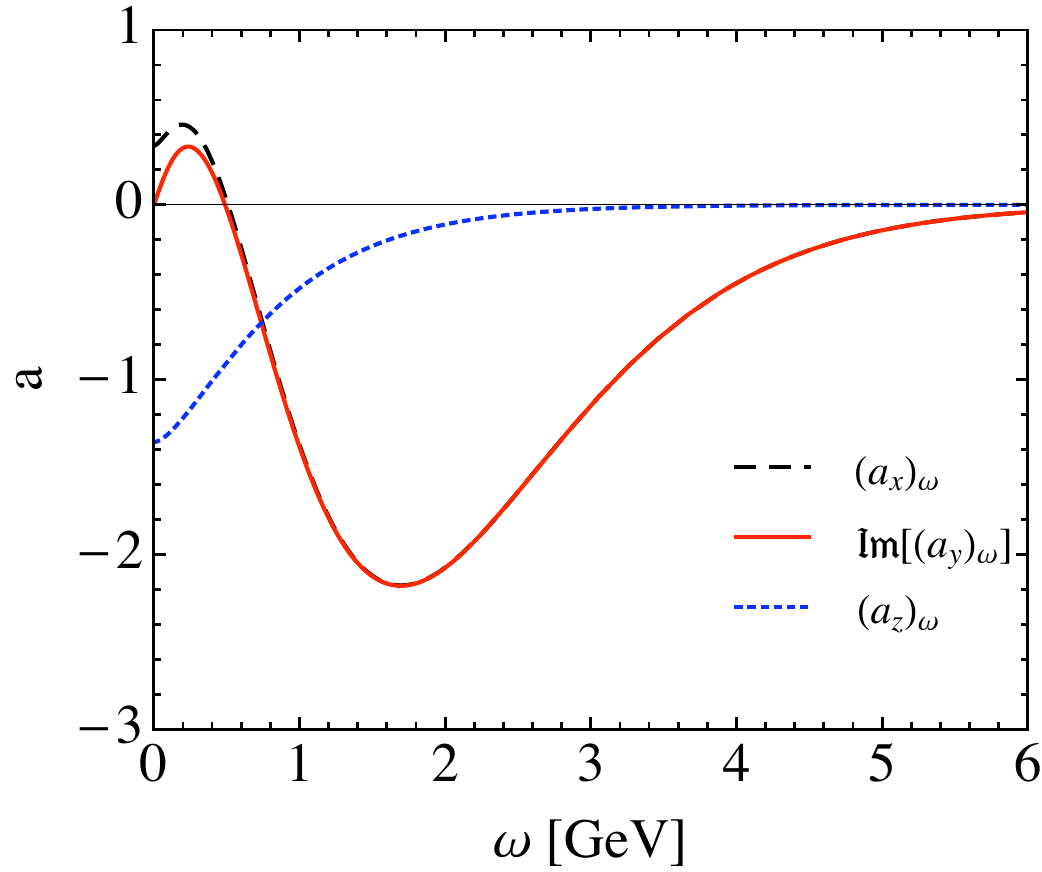}\\}}
\centerline{(a)}
\end{minipage}
\hspace{.24cm}
\begin{minipage}{.48\textwidth}
\parbox{6cm}{
\scalebox{.73}{
\includegraphics{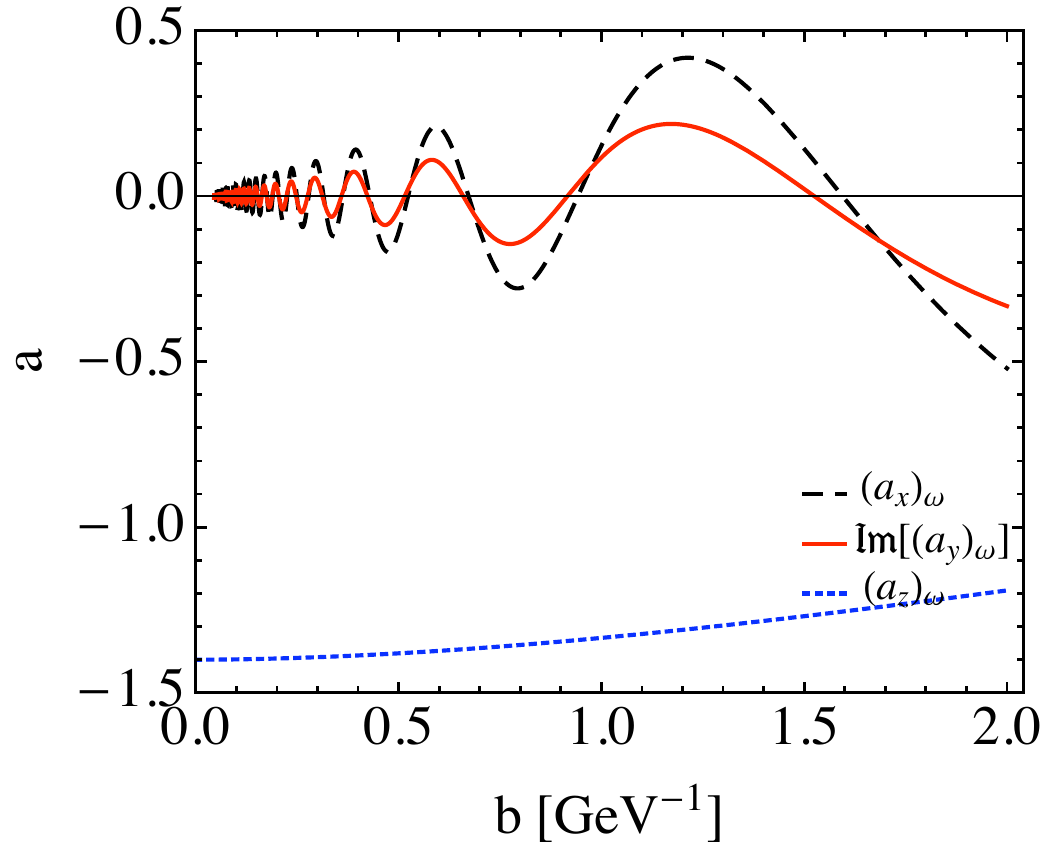}\\}}
\centerline{(b)}
\end{minipage}
\caption{(a): Fourier components of the acceleration as functions of $\omega$, for $m=0.3$ GeV, $b=1$ GeV$^{-1}$, $v=0.7$. (b): Fourier components of the acceleration as functions of $b$, for $m=0.3$ GeV, $\omega=0.1$ GeV, $v=0.7$.}
\label{a}
\end{figure}

\section{Application to QGP}
In this Section we give a rough estimate of the effect that we are describing, in the case in which
the electric charge is a quark $q$ (or an antiquark $\bar{q}$), scattering on a color-magnetic monopole in a deconfined medium. The medium contains a finite density of quarks and
monopoles. We will compare our result to the radiation produced from Coulomb scattering of 
$q\bar{q}$, $\bar{q}\bar{q}$ and $qq$ pairs. We consider a regime of temperatures $\gsim$ 2 $T_c$ where, according to the magnetic scenario proposed in \cite{Liao:2006ry}, monopoles can be considered as heavy, static particles.

A quark moving in a deconfined medium acquires a thermal mass due to its interaction with
the other particles of the medium. Lattice results for this quantity are available for three values of
the temperature~\cite{Karsch:2007wc,Karsch:2009tp}. At $T=1.5T_c$ and $T=3T_c$
they find $m=0.8 T$, while at $T=1.25 T_c$ they obtain
$m=0.77 T$. At $T=2T_c$ we therefore assume a value of $m\simeq 0.8T$, namely $m\simeq 0.3$ GeV. 
For temperatures $T\simeq 2T_c$, we can assume that quarks move with an average velocity
$v\sim0.5-0.7$. 

\begin{figure}
\begin{center}
\scalebox{1}{\includegraphics{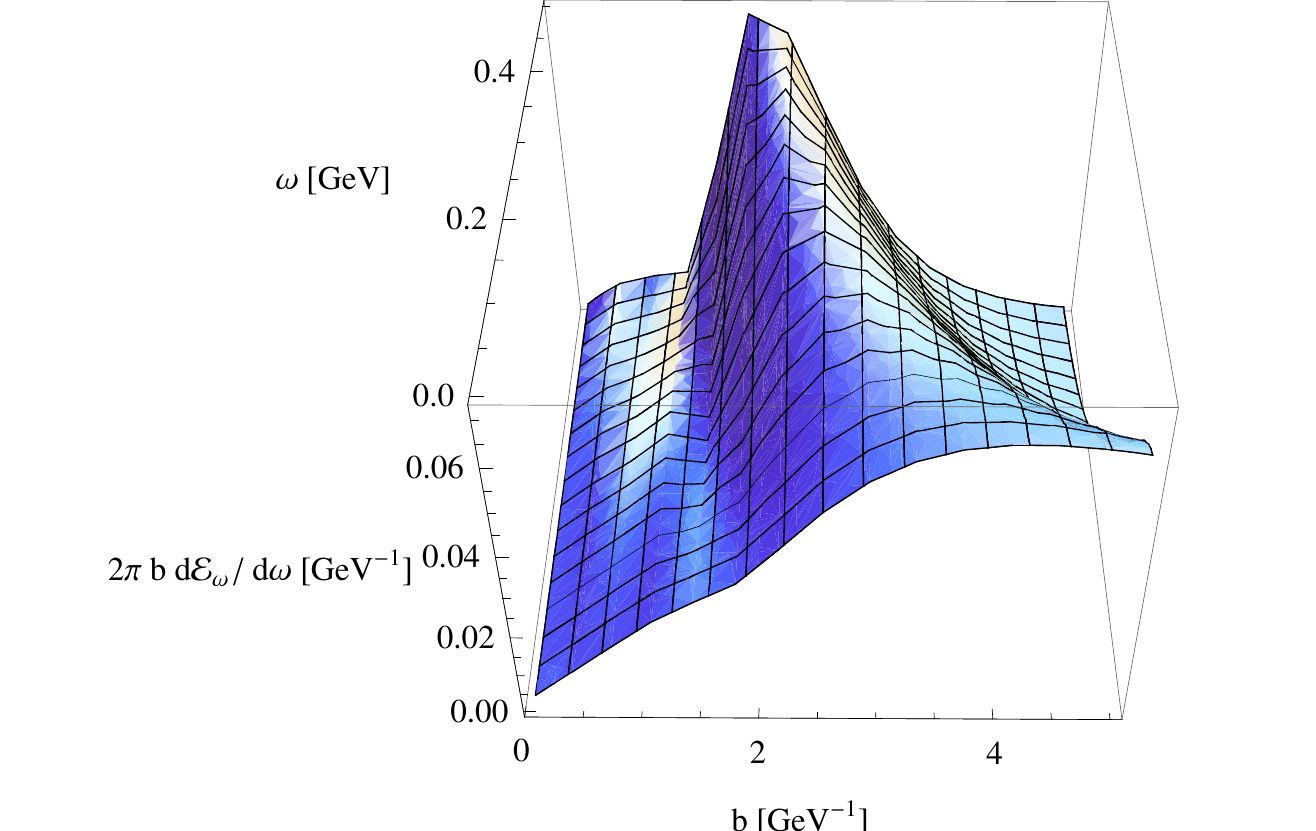}}
\caption{Integrand $2\pi bd\mathcal{E}_\omega/d\omega$ as a
function of $\omega$ and $b$. In this figure, $m=0.3$ GeV and $v=0.7$.
}
\label{fig3}
\end{center}
\end{figure}

In Figure \ref{fig3} we show the integrand for the total radiation in a given frequency interval, namely  $2\pi bd\mathcal{E}_\omega/d\omega$, as a function of $\omega$ and $b$.
We need to integrate this quantity over the impact parameter. There is no upper limit on $b$ in the case of scattering of one quark on one monopole.
However, in matter there is a finite density of
 monopoles, so the scattering issue should be
reconsidered.
 A sketch of the setting,
assuming strong correlation of monopoles into a crystal-like
structure, is shown
 in Fig. \ref{fig_monos_2d}.  A
 ``sphere of influence of one monopole''(the dotted circle)
gives the maximal impact parameter
to be used
\beq 
b_{max}=n_{M}^{-1/3}/2.
\eeq
\begin{figure}[t]
\begin{center}
\includegraphics[width=4cm]{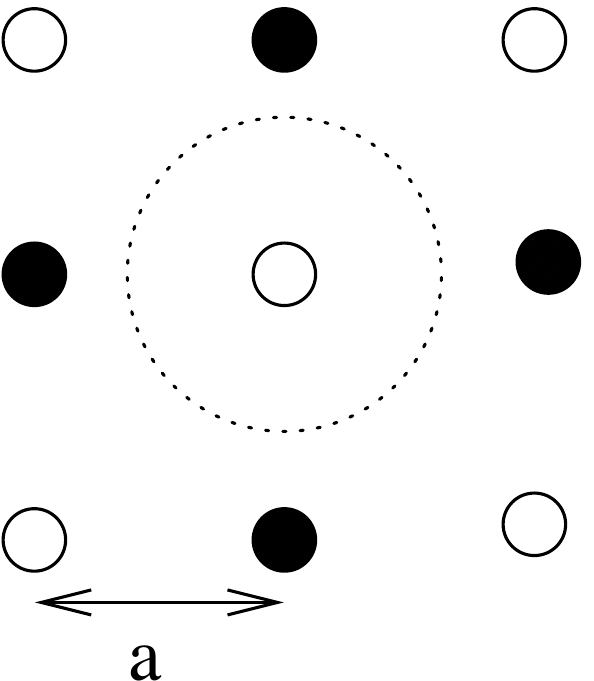}
\caption{A charge scattering on a 2-dimensional array of 
correlated monopoles (open points) and antimonopoles (closed points).
The dotted circle indicates a region of impact parameters for
which scattering on a single monopole is a reasonable approximation.}
\label{fig_monos_2d}
\end{center}
\end{figure}

The same is true for the quark-quark and quark-antiquark Coulomb scattering to which we
will compare our results.
The monopole density as a function of the temperature for $SU(2)$ gauge theory has been
evaluated on the lattice \cite{D'Alessandro:2007su}. In order to account for the transition from 
$SU(2)$ to $SU(3)$ gauge group, we scale these results by a factor 2: in $SU(2)$
there is in fact one monopole species while in $SU(3)$ there are two, identified by two different $U(1)$ subgroups. The monopole
density at $T\simeq2T_c$ is $n_M\simeq 0.02$ GeV$^3$, which gives
$b_{max}\simeq1.8$ GeV$^{-1}$.
The lattice also gives us information about the monopole size
\cite{Ilgenfritz:2007ua}: they turn out to be
very small objects, having a radius $r_M\simeq0.15$ fm=0.78 GeV$^{-1}$.
This is the $b_{min}$ that we will use in our integration.

We therefore obtain, for the total energy radiated in unit volume throughout the time of the
collision in a given frequency interval:
\bea
\frac{d\Sigma}{d\omega}=\frac29\frac{d\kappa_{\omega}}{d\omega}n_qn_M=
\frac29n_qn_M\frac{2}{3\pi}\alpha_{em}2\pi\int_{b_{min}}^{b_{max}}b|\vec{a}_\omega|^2db
\eea
where the factor $\frac29=\frac13(\frac49+\frac19+\frac19)$ comes from the different electric charges
fur $u,~d$ and $s$ quarks (the density of quarks $n_q$ is the sum of the densities of $u,~d$ and $s$ quarks).

\subsection{Comparison with Coulomb scattering}
In this Section we give an estimate of the radiation produced in the scattering of $qq$, $q\bar{q}$
and $\bar{q}\bar{q}$ pairs in the plasma. 
In the case of an attractive interaction between particles (namely in the singlet channel for the
$q\bar{q}$ scattering and the antitriplet channel for the $qq$ and $\bar{q}\bar{q}$ scatterings),
the formula for $d\mathcal{E}_\omega/d\omega$ reads \cite{LL2}:
\bea
\frac{d\mathcal{E}_\omega}{d\omega}=\frac{2\pi\alpha^2\omega^2}{3v^4}\left(\frac{(e_{em})_1}{m_1}-
\frac{(e_{em})_2}{m_2}\right)^2\left\{\left[H_{i\nu}^{(1)\prime}(i\nu\epsilon)\right]^2
+\frac{\epsilon^2-1}{\epsilon^2}\left|H_{i\nu}^{(1)}(i\nu\epsilon)\right|^2\right\}
\label{dedomega}
\eea
where:
\beq
\nu=\frac{\omega\alpha}{\mu v^3},~~~~~~~~~\epsilon=\sqrt{1+\frac{\mu^2b^2v^4}{\alpha^2}},~~~~~~~~~\mu=\frac{m_1m_2}{m_1+m_2}
\eeq
and $H_{i\nu}^{(1)}(i\nu\epsilon)$ is the Hankel function of the first kind:
\beq
H_{n}^{(1)}(z)=J_{n}(z)+iY_{n}(z).
\eeq
In Eq. (\ref{dedomega}), $(e_{em})_1,~m_1$ and $(e_{em})_2,~m_2$ are the electric charge and mass
of the two colliding particles, while $\alpha=C^R\,\alpha_s$ 
is the strong coupling constant multiplied by the corresponding Casimir factor $C^R$ for the channel under study:
$$
C^8\,=\,4/3\,; \ \ \ \ \ \ \ \ C^1\,=\,1/6\,;\ \ \ \ \ \  \ \ \ C^6\,=\,2/3\,;\ \ \ \ \ \ \ \ \ \ C^{\bar 3}\,=\,1/3\,. 
$$
When the interaction is repulsive (namely in the octet channel for the $q\bar{q}$ scattering, and 
in the sextet channel for the $qq$ and $\bar{q}\bar{q}$ scatterings), Eq. (\ref{dedomega})
gets modified as follows:
\bea
\frac{d\mathcal{E}_\omega}{d\omega}&=&\frac{2\pi\alpha^2\omega^2}{3v^4}\left(\frac{(e_{em})_1}{m_1}-
\frac{(e_{em})_2}{m_2}\right)^2\!\left\{\left[H_{i\nu}^{(1)\prime}(i\nu\epsilon)\right]^2\!
\right.
\nonumber\\
~~~~~~~~~~~~~~~~~~&+&\left.
\frac{\epsilon^2\!-\!1}{\epsilon^2}\left|H_{i\nu}^{(1)}(i\nu\epsilon)\right|^2\right\}\exp\left[-2
\pi\nu\right].
\label{dedomegarep}
\eea
We consider quark matter with three equal mass light flavors, namely
$m_1=m_2=m;~\mu=m/2$. The total density of quarks and antiquarks can be
obtained for example from the PNJL model (see Appendix B).
$$
n_q\simeq2.8T^3\simeq0.12~\mathrm{GeV}^3.
$$
The total energy radiated  in unit volume throughout  
the time of the collision in the case of Coulomb scattering can be obtained from the following formula:
\bea
\frac{d\Sigma}{d\omega}\!\!\!\!\!&=&\!\!\!\!\!\frac{4\pi^2\alpha_{em}\omega^2}{3v^4}\frac{n_{q}^{2}}{18^2}
\frac{4}{m^2}\int_{0}^{b_{max}}\!\!\left[\left(\frac43\alpha_s\right)^2\!\!\!f\left(\frac43\alpha_s\right)+8
\left(\frac16\alpha_s\right)^2\!\!\!f\left(\frac16\alpha_s\right)\exp[-2\pi\nu\alpha_s/6]\right.
\nonumber\\
&+&
\left.3\left(\frac23\alpha_s\right)^2f\left(\frac23\alpha_s\right)+6
\left(\frac13\alpha_s\right)^2f\left(\frac13\alpha_s\right)\exp[-2\pi\nu\alpha_s/3]
\right]bdb,
\label{dSigma}
\eea
where
\beq
f(\alpha)=
\left[H_{i\nu}^{(1)\prime}(i\nu\epsilon)\right]^2+
\frac{\epsilon^2-1}{\epsilon^2}\left|H_{i\nu}^{(1)}(i\nu\epsilon)\right|^2.
\eeq
Eq.~(\ref{dSigma}) is the total radiated energy, it takes into account all possible 
color channels for $qq$, $q\bar{q}$ and $\bar{q}\bar{q}$ scatterings, and all possible flavor combinations. $b_{max}$ can be estimated directly from the quark density: in our temperature regime, it turns out that $b_{max}\simeq1$ GeV$^{-1}$.
Our results for the ratio of the total energy radiated throughout the collision time in the case of quark-quark and quark-monopole scattering are shown in Figs. \ref{Sigma}. The left panels show this ratio for $b_{max}\rightarrow\infty$, 
while in the right panels $b_{max}$ is finite and fixed by the corresponding densities. 
This is useful to understand how strongly a finite $b_{max}$ influences our results: it turns out that the cutoff dependence of $d\Sigma/d\omega$ is dramatic. Without cutoff, this quantity is much larger in the
case of a Coulomb scattering, while the opposite is true in the case of a finite $b_{max}$. This effect is qualitatively true for all values of $v$ that we have considered, but obviously the
relative magnitude for $qq$ and $qM$ scatterings depends on the specific value of $v$ that we choose.

\begin{figure}
\hspace{-.8cm}
\begin{minipage}{.48\textwidth}
\parbox{6cm}{
\scalebox{.6}{
\includegraphics{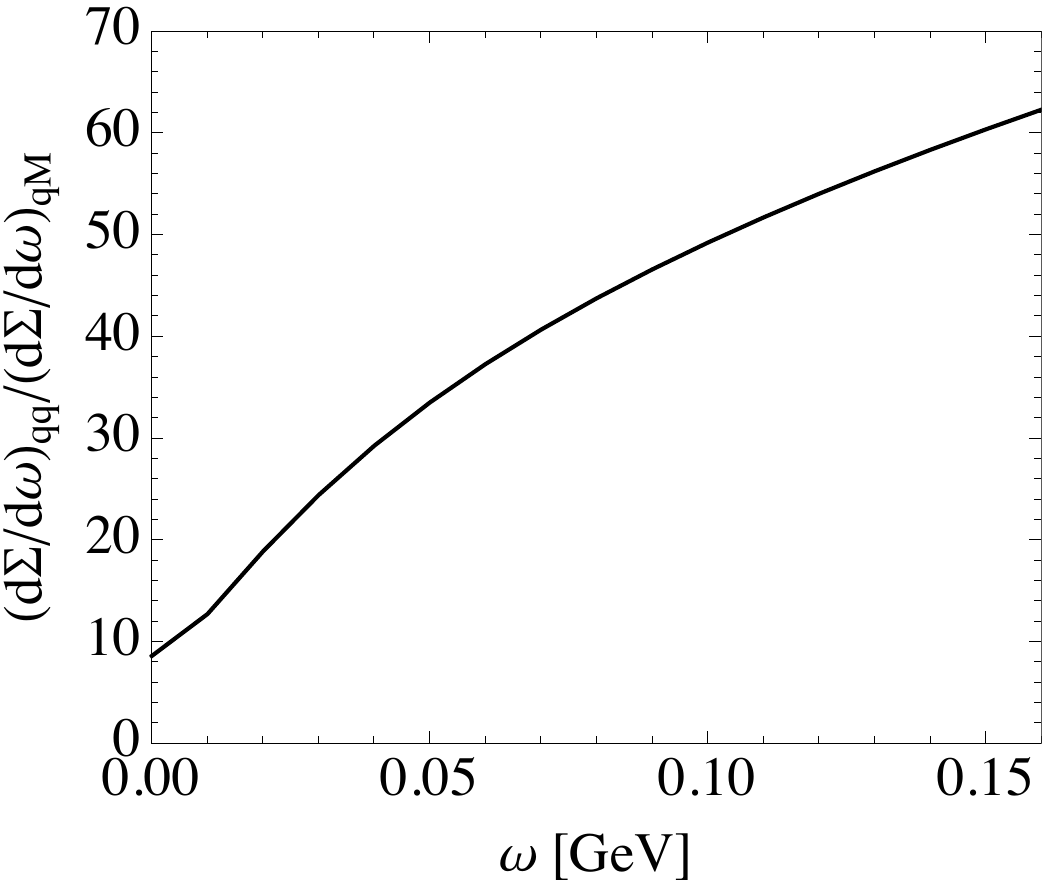}\\}}
\centerline{(a)}
\end{minipage}
\hspace{.2cm}
\begin{minipage}{.48\textwidth}
\parbox{6cm}{
\scalebox{.6}{
\includegraphics{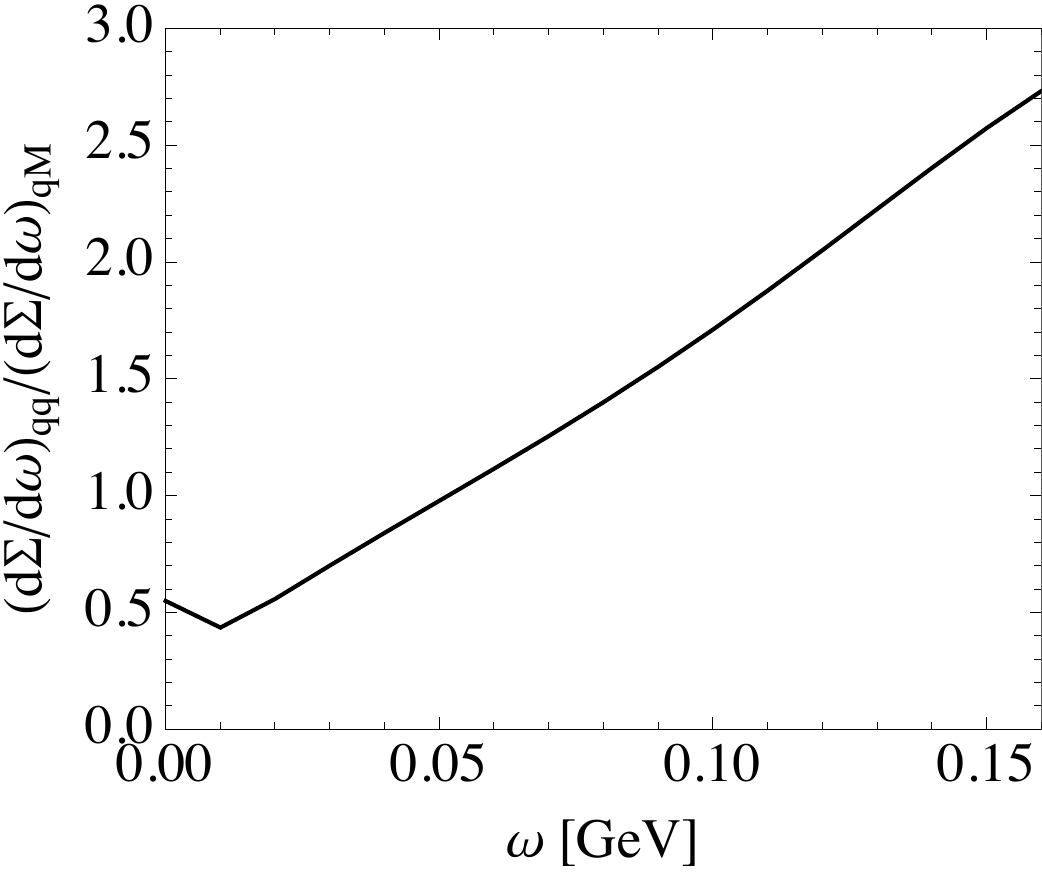}\\}}
\centerline{(b)}
\end{minipage}
\begin{minipage}{.48\textwidth}
\parbox{6cm}{
\scalebox{.6}{
\includegraphics{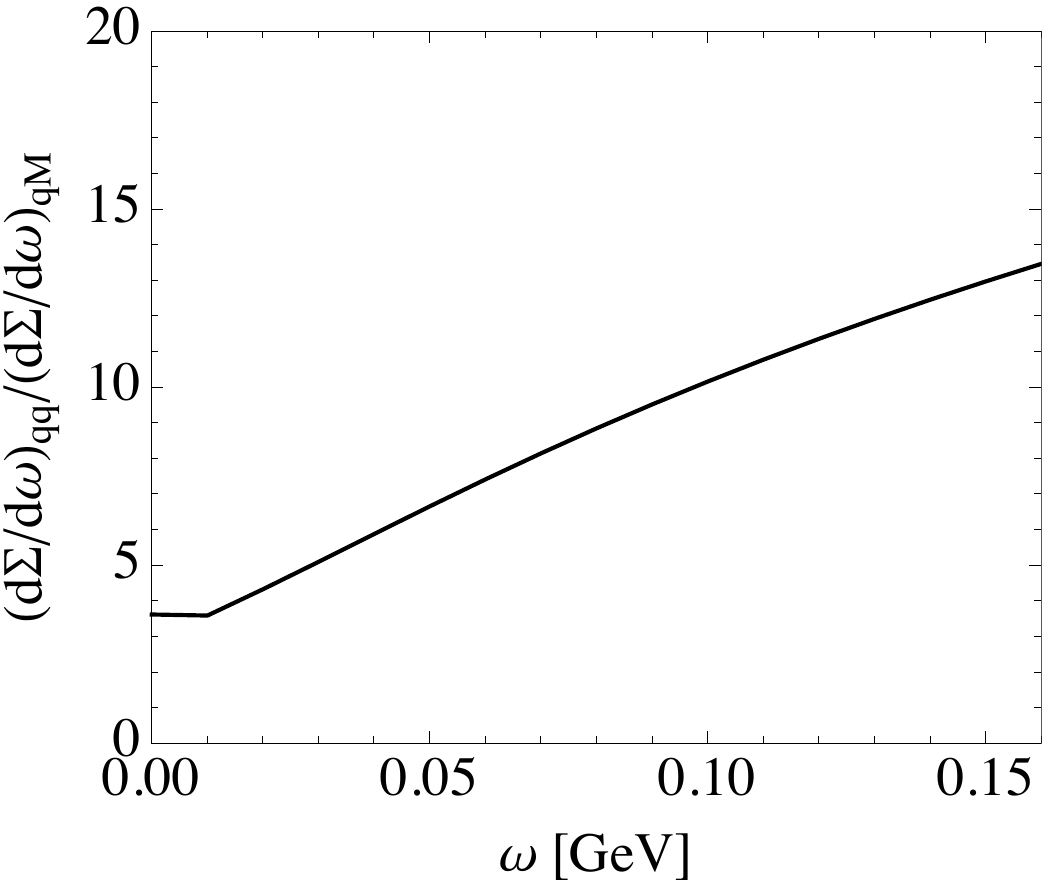}\\}}
\centerline{(c)}
\end{minipage}
\hspace{.2cm}
\begin{minipage}{.48\textwidth}
\parbox{6cm}{
\scalebox{.6}{
\includegraphics{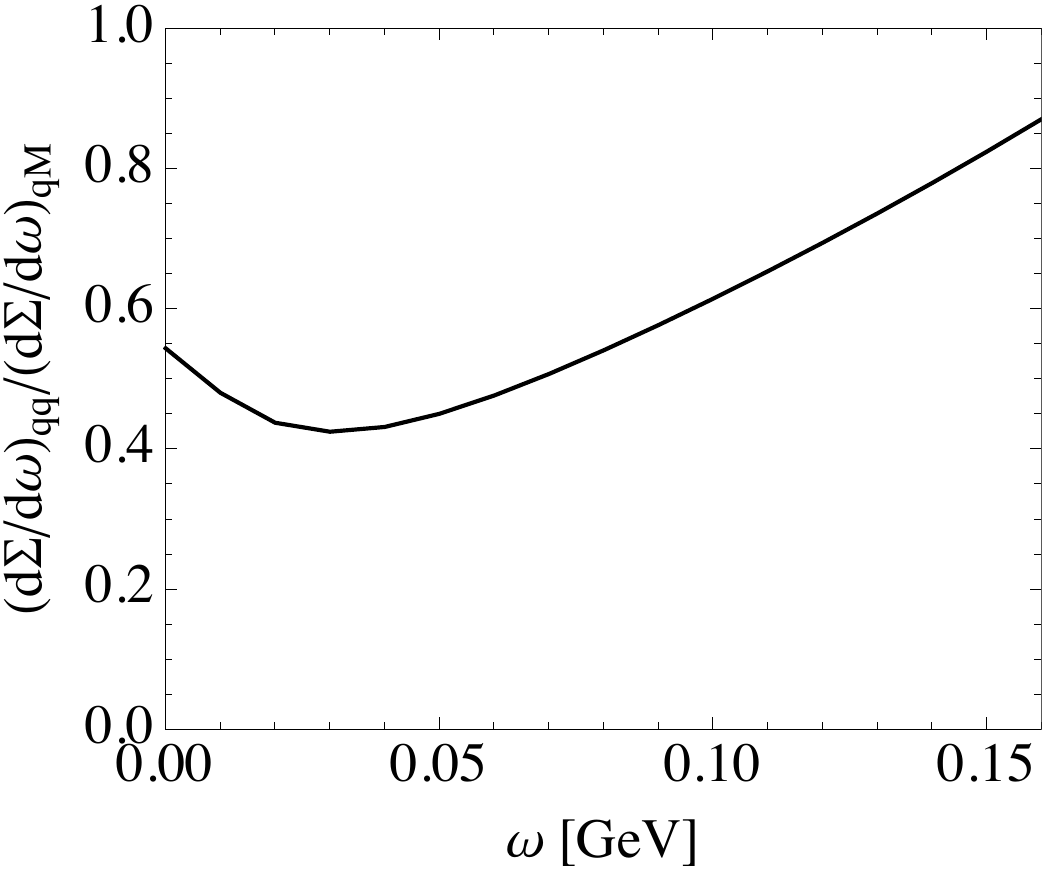}\\}}
\centerline{(d)}
\end{minipage}
\begin{minipage}{.48\textwidth}
\parbox{6cm}{
\scalebox{.6}{
\includegraphics{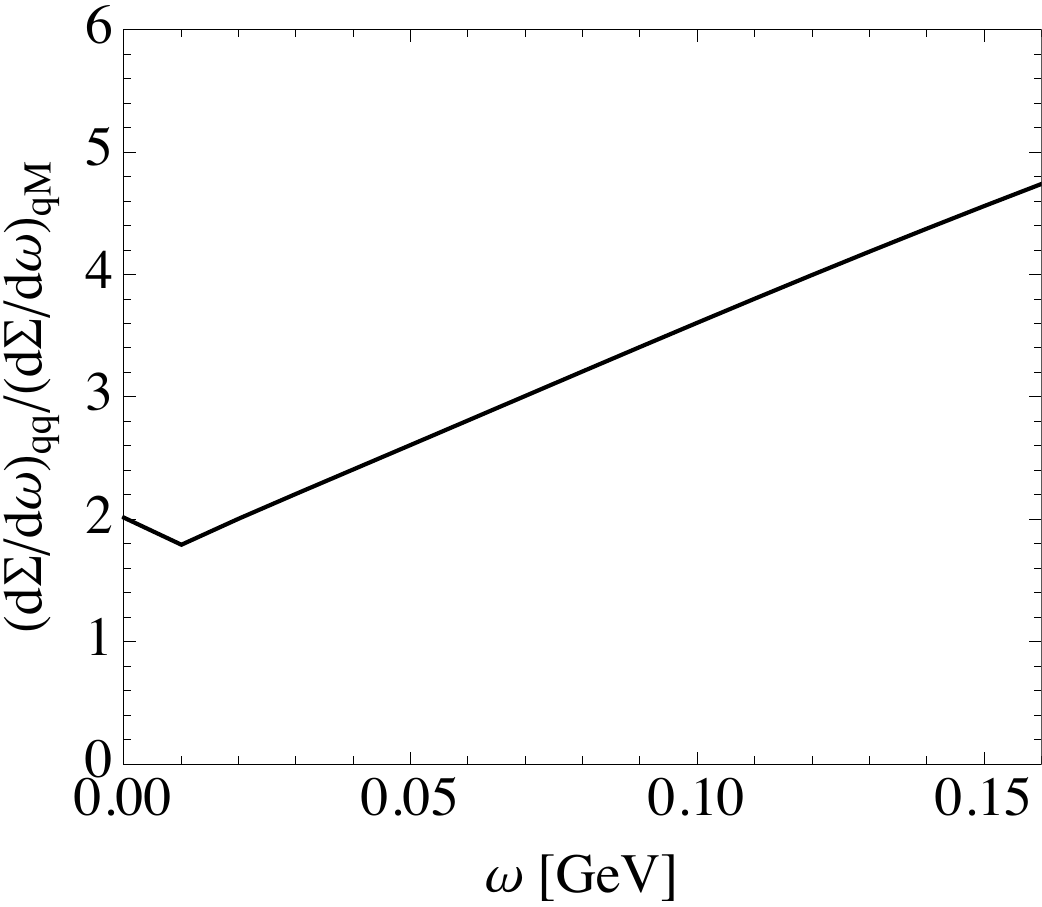}\\}}
\centerline{(e)}
\end{minipage}
\hspace{.2cm}
\begin{minipage}{.48\textwidth}
\parbox{6cm}{
\scalebox{.6}{
\includegraphics{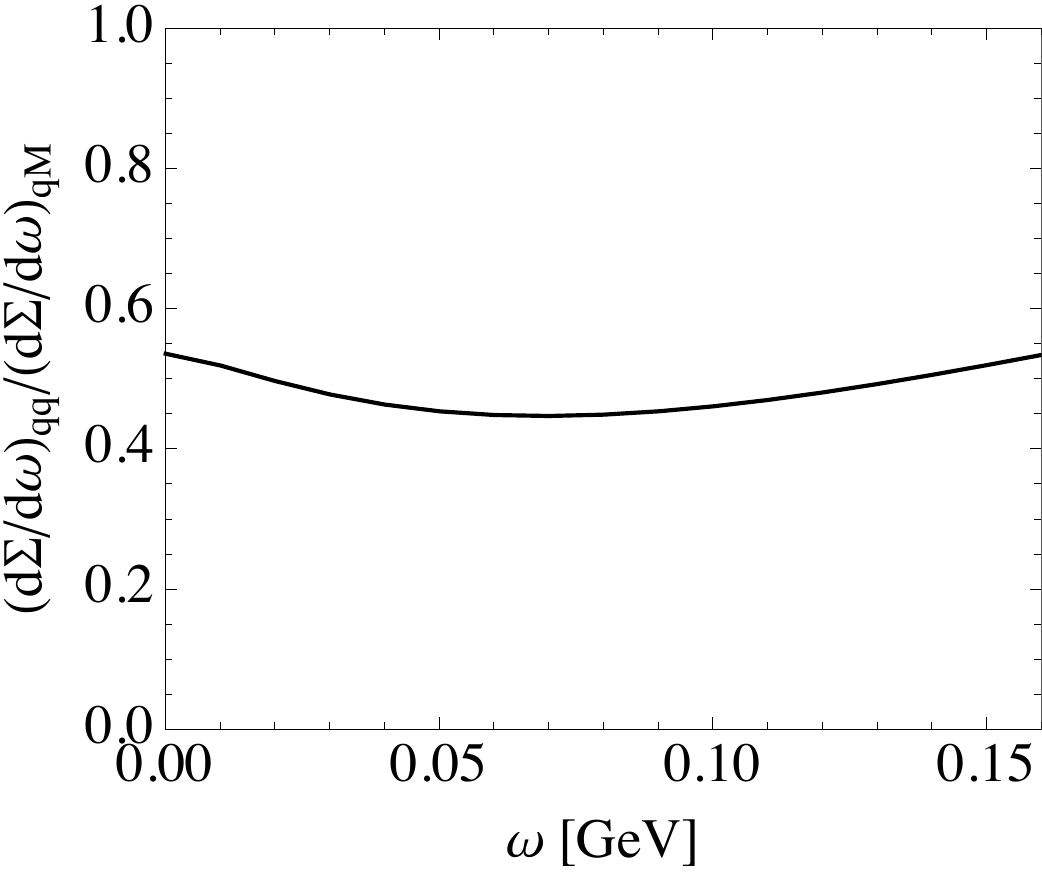}\\}}
\centerline{(f)}
\end{minipage}
\caption{Left column: ratio of $d\Sigma/d\omega$ for Coulomb and quark-monopole
scattering. In both cases, the integral over $b$ is taken up to $\infty$. For these plots we use
$\alpha_s=0.8,~m=0.3$ GeV and: (a) $v=0.3$, (c) $v=0.5$, (e) $v=0.7$. Right column: same as
in the left column, but the integral over $b$ is taken up to the
corresponding $b_{max}$. For these plots we use
$\alpha_s=0.8,~m=0.3$ GeV and: (b) $v=0.3$, (d) $v=0.5$, (f) $v=0.7$.}
\label{Sigma}
\end{figure}
\subsection{Validity of our approximations}
Our results are obtained through a series of approximations:
\begin{itemize}
\item{non-relativistic approximation: this is obviously valid if the velocity $v$ of quarks 
is not too large compared to the speed of light, namely $v\ll1$;}
\item{dipole approximation vs retarded emission. It is valid if the radiation wavelength is large. The retardation effects can be neglected in cases
where the distribution of charge changes little during the time $a/c$, where $a$ is the order of magnitude of the dimensions of the system. This is true if
$v\ll1$, which coincides with the condition for the non-relativistic approximation;}
\item{classical trajectory is used, without back reaction of radiation: in all our formulas, we assume that the particle
moves along a trajectory which is the solution of the classical equations of motion. This means
that 
the energy lost by the particle through the radiation
process is negligible. This approximation is valid in the limit $\omega\ll m/e^2$;}
\item{classical approximation: naively, the emission of soft photons can be described classically for

$$\hbar \,\omega\,\ll E_k\,=\,\mu\,v^2/2\,=\,m\,v^2/4.$$Within the same approximation we can ignore the recoil effect (energy-momentum conservation).

The full quantum treatment of the radiation should include the back reaction of the radiation, 
 one has to evaluate the $nondiagonal$ matrix element of the dipole moment between the initial and final scattering states, with
different energies. Such quantum states for quark-monopole problem were found in \cite{Kazama:1976fm}
and recently for  gluon-monopole problem  in \cite{RS}.
However, the matrix elements have not been computed yet. Since it was done for quantum Coulomb scattering, by A.Sommerfield in 1931,  we can use 
those results in order to have at least some qualitative estimate for the accuracy of  classical description. 

We compare the total radiation $d\kappa_\omega$ in a given frequency interval $d\omega$,
in the case of scattering on a single particle, namely we
integrate over the impact parameter $b$ from $0$ to $\infty$.
In the classical case we have:
\beq
\frac{d\kappa_\omega}{d\omega}=\frac{32\pi^2\omega\alpha_{em}\alpha_s^3}{3m^3v^5}
\left|H_{i\nu}^{(1)}(i\nu)\right|H_{i\nu}^{(1)\prime}(i\nu)
\eeq
which is to be compared to the quantum expression \cite{LL4}:
\begin{equation}
\frac{d\,\kappa(\omega)}{d\omega}\,=\,\alpha_{em}\,\alpha_s^2
\,{64\,\pi^2\over 3}\,{p^\prime\over p}\,
{1\over(p\,-\,p^\prime)^2}\,{1\over(1\,-\,e^{-2\,\pi\,\beta^\prime})\,
(e^{2\,\pi\,\beta}-\,1)}\,\left(-{d\over d\xi}\,|F(\xi)|^2\right)
\end{equation}

where
$$\beta\,=\,{\alpha_s\over\,v}\,\ \ \ \ \ \ \ \ \ \
\beta^\prime\,=\,{\alpha_s\over\,v^\prime}\,\ \ \ \ \ \ \ \ \ \
m\,v\,=2\,p\,\ \ \ \ \ \ \ \ \ \
m\,v^\prime\,=2\,p^\prime\,\ \ \ \ \ \ \ \ \ \
p^\prime\,=\,\sqrt{p^2\,-\,m\,\omega}
$$
and
$$
F(\xi)\,\equiv\,_2F_1(i\,\beta^\prime,\,i\,\beta;\,1;\,\xi)\,\ \ \ \ \ \ \ \ \ \ \ \ \ 
\xi\,=\,-\,{4\,p\,p^\prime\over (p\,-\,p^\prime)^2}
$$
$_2F_1(a,b;c;d)$ is the Hypergeometric  function.
The two curves corresponding to classical and quantum scatterings are
shown in Fig.~\ref{quantum}. 
It is thus evident that within the energy region plotted the classical result is a very good approximation
of the quantum one. Note that there is a maximum $\omega$, which is equal to the energy of the incoming particle
beyond which due to energy conservation the quantum formula are not applicable. 
}

\item{two body vs many body scattering. So far, we have limited our analysis to two-body scattering, while mimicking
the effects of multiple interactions by finite quark and monopole densities and the maximal impact parameter. The strong dependence 
on the maximal impact parameter cutoff observed in our results, reflects  additional shortcomings of our approach. 
The origin of this problem is obvious: on the one hand soft radiation is emitted from large distances; on the other
hand, too large distances are precisely governed by multiple scattering.
}

\end{itemize}
\begin{figure}
\begin{center}
\scalebox{.85}{\includegraphics{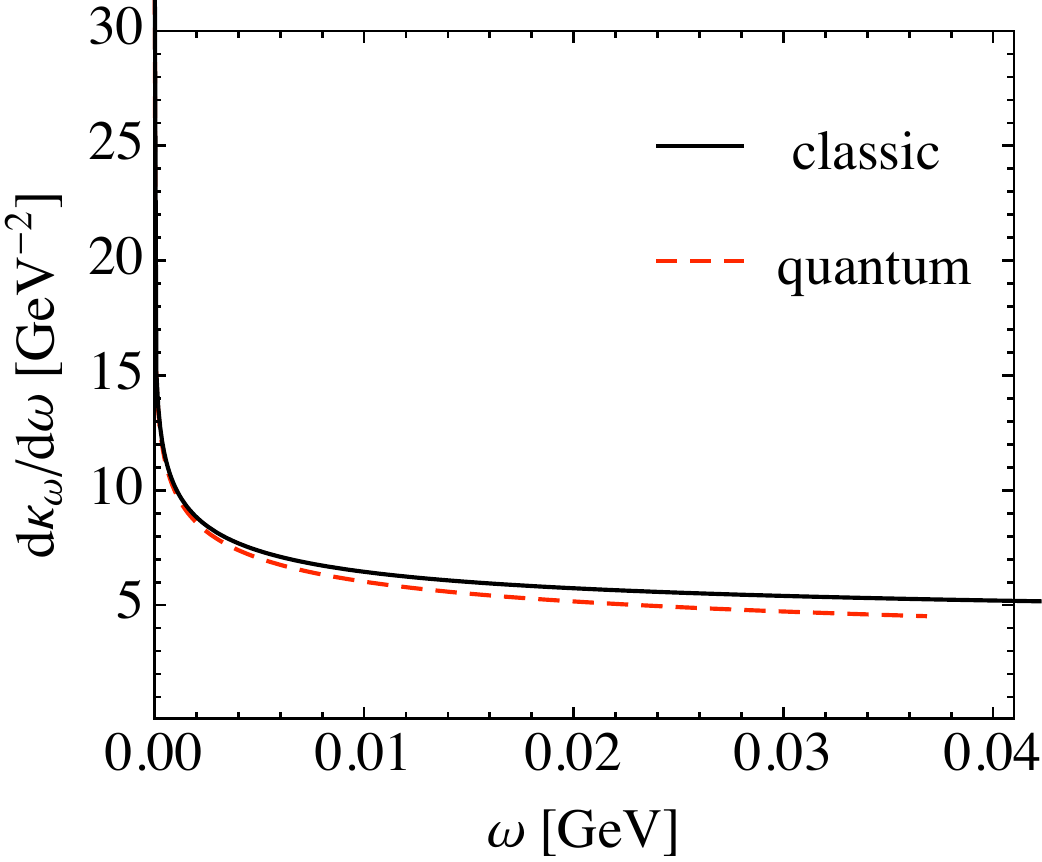}}
\caption{Comparison between classic and quantum result for $d\kappa_\omega/d\omega$
as a function of $\omega$, with
$\alpha_s=0.8,~v=0.7,~m=0.3$ GeV.
}
\label{quantum}
\end{center}
\end{figure}

\section{Conclusions and outlook}
The present paper is the first step towards understanding whether the contribution of the quark-monopole scattering process in QGP is or is not 
 important for photon and dilepton production. 
Our purpose at present was pretty modest: we wanted to evaluate the magnitude of soft
photon radiation from this process and compare it to  the one
produced in Coulomb quark-quark scattering.  A qualitative estimate outlined in the Introduction, said that while the monopole
density is small at high $T$, $n_m\sim (1/\ln(T))^3$, the square of $\alpha_s$ in the electric scattering cross section compensates
two out of three such logarithms.  So, parametrically,
the process we consider is subleading at very large $\ln(T)$.
On the other hand, the charge-monopole scattering at large angles and small impact parameter tends to be  
much larger than the charge-charge one: we found that this enhances radiation, similarly to how it worked for transport processes \cite{RS}.
This represents an encouraging starting point for future work.

We have calculated the photon radiation rate for quarks scattering on monopoles
in a thermal medium which contains a finite density of both particles.
We worked in the classic, non-relativistic approximation and neglected retardation
effects and back reaction. Therefore,  our calculation has a
rather methodological status, it cannot  address the actual phenomenological
questions, 
such as the experimentally observed excess in dilepton production at small $p_t$ and invariant mass below $m_\rho$.
 We need to improve the present paper in many directions: first of all, a full quantum and
relativistic calculation will be performed, also taking into account back reaction of the radiation.
This can in principle be done using non-diagonal matrix elements, calculated between quantum scattering states such as those which 
were found in \cite{RS,Kazama:1976fm}.
 
Then we need to take into
account the fireball evolution, in order to be able to quantitatively compare our results to
the experimental data. Dramatic expansion of the fireball and long duration of the near-$T_c$ phase leads to the conclusion that soft dileptons currently constituting
the puzzle come from the end of the evolution of QGP, $T\sim 1T_c$, rather than the beginning of it, in the temperature regime we discussed above.  
Unfortunately, the near-$T_c$ region is very complicated and quite challenging theoretically. In this region monopoles become
lighter and thus dynamical and relativistic,  matter is no longer an electric near-perturbative plasma with at least well-defined counting rules,
 but a strongly coupled liquid made of all kind
of quasiparticles.
We hope to address those issues elsewhere in our future work. 

\section*{Acknowledgments}

\noindent This work is partially supported by the DOE grants DE-FG02-88ER40388 and DE-FG03-97ER4014 and by the DFG grant SFB-TR/55.

\section*{Appendix A
\label{appendixa}}
In this Appendix we explicitly calculate the Fourier transform of the $\vec{a}$ components.
We start from $(a_x)$:
\bea
(a_x)_\omega=-\frac{(eg)^2}{m^2b^2v\xi}\int_{-\infty}^{\infty}\frac{\exp\left[i \bar{\omega} t\right]\cos\left[\xi\mathrm{arctan} t\right]}{\left(t^2+1\right)^{3/2}}dt.
\eea
If we set $t=i+i\tau$, $dt=id\tau$. 
Both the square root and the $\mathrm{arctan}$ have branch cut singularities 
from $\tau=0$ to $\tau=\infty$ and from $\tau=-\infty$ to $\tau=-2$. 
This is clear if we write the relationship between $\mathrm{arctan}$ and $\mathrm{log}$:
\beq
\mathrm{arctan}\left(i+i\tau\right)=i\mathrm{arctanh}\left(1+\tau\right)=i\log\left(-\frac{2+\tau}{\tau}\right).
\eeq
We calculate the integral along the contour $C$ shown in Fig. \ref{contour}.
\begin{figure}
\begin{center}
\begin{minipage}{.48\textwidth}
\parbox{6cm}{
\scalebox{.6}{
\includegraphics{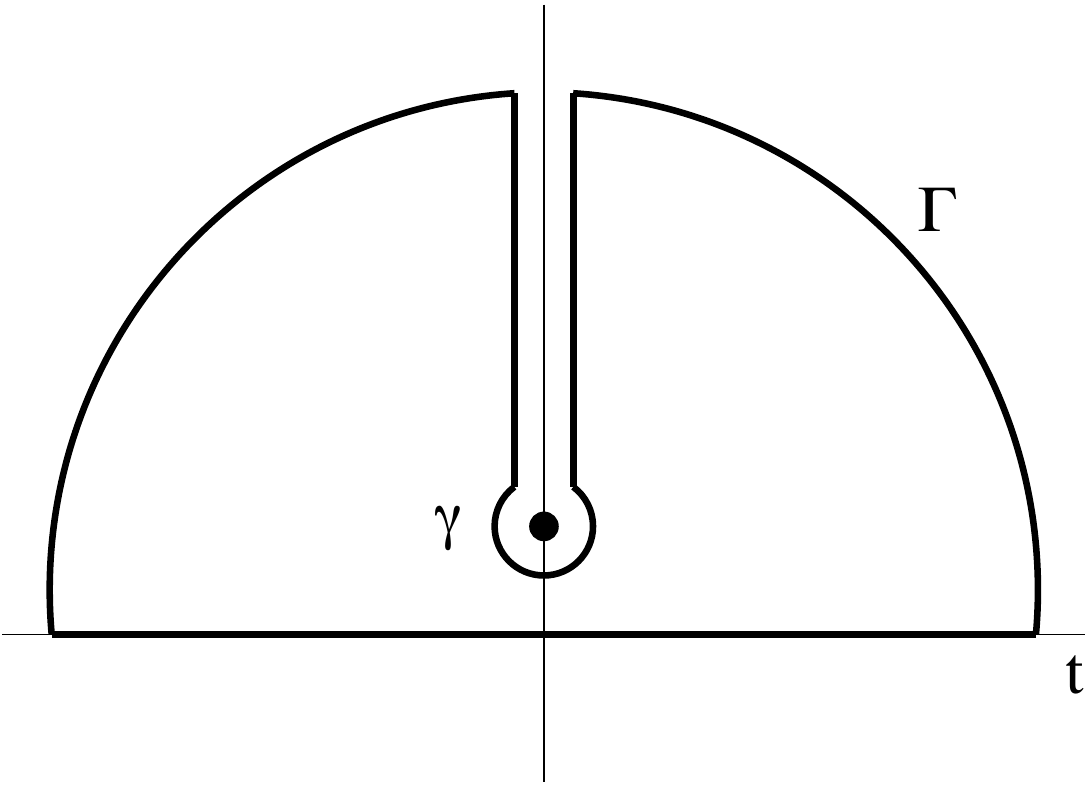}\\}}
\end{minipage}
\begin{minipage}{.48\textwidth}
\parbox{6cm}{
\scalebox{.6}{
\includegraphics{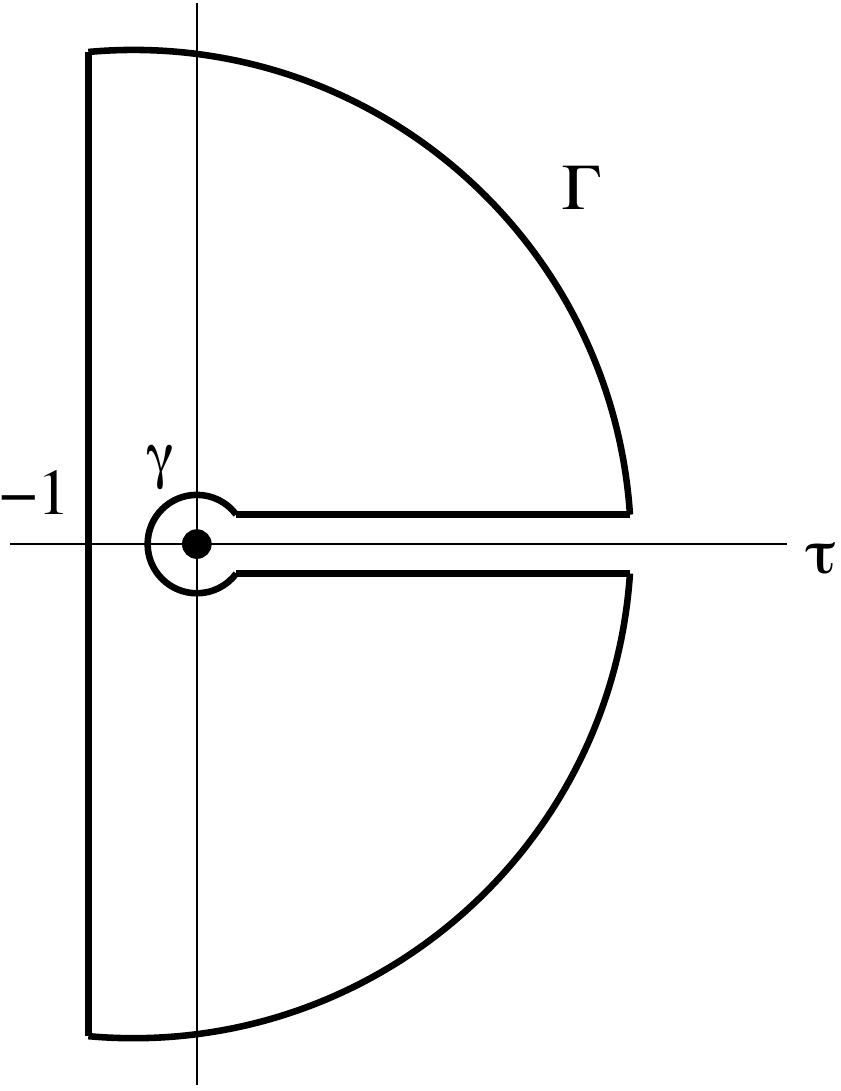}\\}}
\end{minipage}
\caption{Integration contours in the complex $t$ and $\tau$ planes.
}
\label{contour}
\end{center}
\end{figure}
The integral can be split into four different contributions:
\begin{itemize}
\item{the small circle $\gamma$ of radius $\epsilon$}
\item{the upper and lower segments around the branch cut}
\item{the big circle $\Gamma$}
\end{itemize}
We have
\bea
&&\!\!\!\!\!\!\!\!\!\!\!\!\!\!\!\!\!\!\!\!\!\!\!\!\!\!\!\!\!\!\int_{-\infty}^{\infty}\frac{\exp\left[i \bar{\omega} 
t\right]\cos\left[\xi\mathrm{arctan} t\right]}{\left(t^2+1\right)^{3/2}}dt=
\nonumber\\
&&=i\exp\left[-\bar{\omega}\right]\left\{\int_{\epsilon}^{\infty}\!\!\!\!\!\!+\!\int_{\Gamma}\!+\!\int^{\epsilon}_{\infty}\!\!\!+\!\int_{\gamma}\right\}\frac{\exp\left[- \bar{\omega} \tau\right]\cos\left[\xi\mathrm{arctan} \left(i+i\tau\right)\right]}{\left(-2\tau-\tau^2\right)^{3/2}}d\tau.
\eea
Along the upper segment, where $\tau\rightarrow\tau+i\epsilon$, we have:
\bea
\left(-2\tau-\tau^2\right)^{3/2}&=&-i\left(2\tau+\tau^2\right)^{3/2}
\nonumber\\
\log\left(-\frac{2+\tau}{\tau}\right)&=&\log\left|\frac{2+\tau}{\tau}\right|+i\pi
\eea
while along the lower one we have
\bea
\left(-2\tau-\tau^2\right)^{3/2}&=&i\left(2\tau+\tau^2\right)^{3/2}
\nonumber\\
\log\left(-\frac{2+\tau}{\tau}\right)&=&\log\left|\frac{2+\tau}{\tau}\right|-i\pi.
\eea
Therefore we can write:
\bea
&&\!\!\!\!\!\!\!\!\!\!\!\!\!\!\!\!\left\{\int_{\epsilon}^{\infty}\!\!\!\!\!\!+\!\int^{\epsilon}_{\infty}\right\}\frac{\exp\left[- \bar{\omega} \tau\right]\cos\left[\xi\mathrm{arctan} \left(i+i\tau\right)\right]}{\left(-2\tau-\tau^2\right)^{3/2}}d\tau=
\\
&&\!\!\!\!\!\!\!\!\!\!\!\!\!\!\!\!\int_{\epsilon}^{\infty}\frac{\exp\left[-\bar{\omega}\tau\right]\cos\left[\frac{i\xi}{2}\left(\log\left|\frac{2+\tau}{\tau}\right|+i\pi\right)\right]}{-i\left(2\tau+\tau^2\right)^{3/2}}d\tau
+\int^{\epsilon}_{\infty}\frac{\exp\left[-\bar{\omega}\tau\right]\cos\left[\frac{i\xi}{2}\left(\log\left|\frac{2+\tau}{\tau}\right|-i\pi\right)\right]}{i\left(2\tau+\tau^2\right)^{3/2}}d\tau
\nonumber\\
&&~~~~~~~~~~~~~~~~~~~~~~~~~~~~~~~~~~~~=2i\cos\left(\frac{\xi\pi}{2}\right)\int_{\epsilon}^{\infty}\frac{\exp\left[-\bar{\omega}\tau\right]\cos\left[\frac{i\xi}{2}\log\left|\frac{2+\tau}{\tau}\right|\right]}{\left(2\tau+\tau^2\right)^{3/2}}d\tau
\nonumber
\eea
where we used the relationship
\beq
\cos\alpha+\cos\beta=2\cos\frac{\alpha+\beta}{2}\cos\frac{\alpha-\beta}{2}.
\eeq
We therefore have:
\bea
\left\{\int_{\epsilon}^{\infty}\!\!\!\!\!\!+\!\int^{\epsilon}_{\infty}\right\}&&\!\!\!\!\!\!\frac{\exp\left[- \bar{\omega} \tau\right]\cos\left[\xi\mathrm{arctan} \left(i+i\tau\right)\right]}{\left(-2\tau-\tau^2\right)^{3/2}}d\tau
=
\\
&&=i\cos\left(\frac{\xi\pi}{2}\right)\int_{\epsilon}^{\infty}\frac{\exp\left[-\bar{\omega}\tau\right]}{\left(2\tau+\tau^2\right)^{3/2}}\left[\left(\frac{2+\tau}{\tau}\right)^{-\xi/2}+\left(\frac{2+\tau}{\tau}\right)^{\xi/2}\right]d\tau.
\nonumber
\eea
The above integral contains two terms. The first one gives (in the limit $\epsilon\rightarrow0$):
\bea
&&\!\!\!\!\!\!\!\!\!\!\!\!\!\!\!\!\!\!\!\!\!\!\!\!\!\!\!\!\!\!\!\!\!\!\!\!\!\!\!\!\!\!i\cos\left(\frac{\xi\pi}{2}\right)\int_{0}^{\infty}\frac{\exp\left[-\bar{\omega}\tau\right]}{\left(2\tau+\tau^2\right)^{3/2}}\left(\frac{2+\tau}{\tau}\right)^{-\xi/2}\!\!\!\!\!\!d\tau=
\nonumber\\
~~~~~~&&=i\cos\left(\frac{\xi\pi}{2}\right)\frac14\Gamma\left[\frac12\left(\xi-1\right)\right]U\left(\frac12\left(\xi-1\right),-1,2\bar{\omega}\right).
\eea
The second term needs to be integrated by parts $p$ times, where $p$ is the smallest
integer number $>\frac{\xi+3}{2}$. All boundary terms are either divergent or vanishing 
as $\epsilon\rightarrow0$. The divergent ones are exactly cancelled by corresponding divergent terms coming from the integral over the small circle $\gamma$. 
The finite part is:
\bea
&&\!\!\!\!\!\!\!\!\!\!\!\!\int_{\epsilon}^{\infty}\frac{\exp\left(-\bar{\omega}\tau\right)\left(2+\tau\right)^{(\xi-3)/2}}{\tau^{(\xi+3)/2}}d\tau=\frac{4\Gamma\left(\frac{\xi+3}{2}-p\right)p!}{\xi^2-1}\sum_{k=0}^{p}\frac{(-\bar{\omega})^k}{k!
(p-k)!\Gamma(\frac{\xi-3}{2}-p+k+1)}\times
\nonumber\\
\nonumber\\
&&\!\!\!\!\!\!\!\!\!\!\!\!\times\int_{0}^{\infty}\exp\left(-\bar{\omega}\tau\right)\left[2+\tau\right]^{(\xi-3)/2-(p-k)}\tau^{p-(\xi+3)/2}d\tau=
\\
\nonumber\\
&&\!\!\!\!\!\!\!\!\!\!\!\!=\frac{4\Gamma\left(\frac{\xi+3}{2}-p\right)p!}{\xi^2-1}\sum_{k=0}^{p}\frac{(-\bar{\omega})^k}{k!
(p-k)!\Gamma(\frac{\xi-3}{2}-p+k+1)}\times
\nonumber\\
\nonumber\\
&&~~~~~~~~~~~~~~~~~~~~~~~~~~~~~~~~~~~~~~~\times2^{k-2}\Gamma\left(p-\frac{\xi+1}{2}\right)U\left(p-\frac{\xi+1}{2},k-1,2\bar{\omega}\right).
\nonumber
\eea
The above contribution vanishes for odd, integer values of $\xi$. The contribution coming
from $\Gamma$ vanishes identically, while we get a nonvanishing contribution
from the integral over the small circle $\gamma$. This contribution is divergent for
noninteger or even $\xi$, exactly cancelling the divergent contribution
coming from the segments along the branch cut. When $\xi$ is an odd integer number we
get a finite contribution. We redefine $t=i+i\epsilon\exp(i\theta)$,
$dt=-\epsilon\exp(i\theta)d\theta$ and get:
\bea
&&\int_{\gamma}\frac{\exp\left[- \bar{\omega} \tau\right]\cos\left[\xi\mathrm{arctan} \left(i+i\tau\right)\right]}{\left(-2\tau-\tau^2\right)^{3/2}}d\tau
\nonumber\\
&&=-\epsilon e^{-\bar{\omega}}
\int_{0}^{2\pi}\frac{e^{i\theta}e^{-\epsilon\bar{\omega} e^{i\theta}}\cos\left[\xi\mathrm{arctan}\left(i+i\epsilon\exp(i\theta)\right)\right]}{\left(-2\epsilon\exp(i\theta)-\epsilon^2\exp(2i\theta)\right)^{3/2}}d\theta
\nonumber\\
&&=
-\epsilon e^{-\bar{\omega}}
\int_{0}^{2\pi}\frac{e^{i\theta}e^{-\epsilon\bar{\omega} e^{i\theta}}}{\left(-2\epsilon\exp(i\theta)-\epsilon^2\exp(2i\theta)\right)^{3/2}}\cos\left[\frac{\xi}{2i}
\log\left(-\frac{\epsilon e^{i\theta}}{2+\epsilon e^{i\theta}}\right)
\right]d\theta
\nonumber\\
&&=\frac{e^{-\bar{\omega}}}{2(-\epsilon)^{1/2}}
\int_{0}^{2\pi}\frac{e^{-i\theta/2}e^{-\epsilon\bar{\omega} e^{i\theta}}}{\left(2+\epsilon\exp(i\theta)\right)^{3/2}}\left[\left(-\frac{\epsilon e^{i\theta}}{2+\epsilon e^{i\theta}}\right)^{\xi/2}+
\left(-\frac{\epsilon e^{i\theta}}{2+\epsilon e^{i\theta}}\right)^{-\xi/2}
\right]d\theta
\nonumber\\
\eea
For $\epsilon\rightarrow 0$, the first term in the parenthesis gives a finite contribution
only for $\xi=1$, a value which is never reached in practical cases, as we will see.
For $\xi>1$ its contribution vanishes identically. The second term gives
\bea
&&\frac{e^{-\bar{\omega}}}{2(-\epsilon)^{1/2}}
\int_{0}^{2\pi}\frac{e^{-i\theta/2}e^{-\epsilon\bar{\omega} e^{i\theta}}}{\left(2+\epsilon e^{i\theta}\right)^{3/2}}
\left(-\frac{\epsilon e^{i\theta}}{2+\epsilon e^{i\theta}}\right)^{-\xi/2}\!\!\!\!\!\!
d\theta=
\nonumber\\
\nonumber\\
&&=\frac{e^{-\bar{\omega}}}{2}\int_{0}^{2\pi}\frac{e^{-2i\theta}e^{-\epsilon\bar{\omega} e^{i\theta}}\left(2 e^{-i\theta}+\epsilon\right)^{(\xi-3)/2}}{(-\epsilon)^{(\xi+1)/2}}d\theta=
\nonumber\\
\nonumber\\
&&=\frac{e^{-\bar{\omega}}}{2(-\epsilon)^{(\xi+1)/2}}\int_{0}^{2\pi}d\theta e^{-2i\theta}\sum_{n=0}^{\infty}
\frac{\omega^n(-\epsilon)^n e^{in\theta}}{n!}
\sum_{k=0}^{(\xi-3)/2}\frac{(2e^{-i\theta})^k\epsilon^{(\xi-3)/2-k}(\frac{\xi-3}{2})!}{k!(\frac{\xi-3}{2}-k)!}=
\nonumber\\
\nonumber\\
&&=\frac{e^{-\bar{\omega}}}{2}\int_{0}^{2\pi}d\theta e^{-2i\theta}\sum_{k=0}^{(\xi-3)/2}
\frac{\omega^{k+2}}{(k+2)!}
\frac{(-1)^{k-(\xi-3)/2}2^ke^{2i\theta}(\frac{\xi-3}{2})!}{k!(\frac{\xi-3}{2}-k)!}=
\nonumber\\
\nonumber\\
&&=
\frac{e^{-\bar{\omega}}}{2}2\pi\sum_{k=0}^{(\xi-3)/2}
\frac{\omega^{k+2}}{(k+2)!}
\frac{(-1)^{k-(\xi-3)/2}2^k(\frac{\xi-3}{2})!}{k!(\frac{\xi-3}{2}-k)!}
\eea
where we have taken into account the fact that the finite contribution from the sum over $n$ comes from $n=k+2$.
The final result for $(a_x)_\omega$ is therefore
\bea
(a_{x})_{\omega}&=&\frac{(eg)^2}{m^2b^2v\xi}\left\{\exp\left(-\bar{\omega}\right)\cos\left(\frac{\pi\xi}{2}\right)
\left[\frac14\Gamma\left(\frac12\left(\xi-1\right)\right)U\left(\frac12\left(\xi-1\right),-1,2\bar{\omega}\right)\right.\right.
\nonumber\\
\nonumber\\
&\!\!\!\!\!\!\!\!\!\!\!\!\!\!\!\!\!\!\!\!\!\!\!\!\!\!\!\!\!\!\!\!\!\!\!\!\!\!\!\!\!\!\!\!\!\!+&\!\!\!\!\!\!\!\!\!\!\!\!\!\!\!\!\!\!\!\!\!\!\!\!\!\!\!\!\left.\left.\frac{4p!\Gamma\left(-p+\frac{\xi+3}{2}\right)}{\xi^2-1}\sum_{k=0}^{p}\frac{\left(-\bar{\omega}\right)^k}{k!\left(p-k\right)!\Gamma\left(\frac{\xi-3}{2}-p+k+1\right)}\right.\right.
\times
\label{wx2}\\
\nonumber\\
&&~~~~~~~~~~~~~~~~~~~~~~\times\left.\left.2^{k-2}
\Gamma\left(p-\frac{\xi+1}{2}\right)U\left(p-\frac{\xi+1}{2},k-1,2\bar{\omega}\right)
\right]\right\}
\nonumber\\
\eea
for any value of $\xi$ other than odd-integer, and
\beq
(a_x)_\omega=\frac{(eg)^2}{m^2b^2v\xi}
\frac{e^{-\bar{\omega}}}{2}2\pi\sum_{k=0}^{(\xi-3)/2}
\frac{\omega^{k+2}}{(k+2)!}
\frac{(-1)^{k-(\xi-3)/2}2^k(\frac{\xi-3}{2})!}{k!(\frac{\xi-3}{2}-k)!}.
\label{oddxi}
\eeq
for odd-integer $\xi$.

The component $(a_y)_\omega$ slightly differs from $(a_x)_\omega$: the $\sin$ integration
gives a purely imaginary contribution. The second term in Eq. (\ref{wx2}) has a minus sign.

The first terms of the components expansion around $\omega=0$ have the following assymptotic
behavior:
\bea
&&\!\!\!\!\!\!\!\!\!\!(a_x)_{\omega\rightarrow0}\simeq\frac{2v}{\xi}\cos\left[\frac{\pi\xi}{2}\right]+\omega^2\frac{b^2(\xi^2-1)}{v\xi}
\cos\left[\frac{\pi\xi}{2}\right]\left\{
-\frac{1}{\xi^2-1}\right.
\nonumber\\
&&\!\!\!\!\!\!\!\!\!\!+\left.\frac12\left[-\mathrm{Harmonicnumber}\left(p-\frac{\xi+3}{2}\right)
-\mathrm{Harmonicnumber}\left(\frac{\xi+1}{2}\right)
\right.\right.
\\
&&\!\!\!\!\!\!\!\!\!\!+\left.\left.
\frac1{\xi^2-1}\left(-3-2p^2-2p(\xi-2)-2\gamma\left(\xi^2-1\right)+
\xi\left(4-\xi\left(-3+\ln4\right)\right)
\right.\right.\right.
\nonumber\\
&&\!\!\!\!\!\!\!\!\!\!+\left.\left.\left.\ln4-2\left(\xi^2-1\right)\ln\frac bv\omega\right)\right]
+\theta\left(p-3\right)\frac{4p!\Gamma\left[-p+\frac{\xi+3}{2}\right]}{\xi^2-1}\sum_{k=3}^{p}\frac{(-1)^k(k-3)!}{k!(p-k)!\Gamma\left[\frac{\xi-3}{2}-p+k+1\right]}
\right\}
\nonumber\\
\nonumber\\
\nonumber\\
&&\!\!\!\!\!\!\!\!\!\!(a_y)_{\omega\rightarrow0}\simeq 2ib\cos\left[\frac{\pi\xi}{2}\right]\omega
\nonumber\\
\nonumber\\
\nonumber\\
&&\!\!\!\!\!\!\!\!\!\!(a_z)_{\omega\rightarrow0}\simeq-\frac{(eg)v}{\sqrt{(eg)^2+(mvb)^2}}
-\omega^2\left[\frac{b^2eg\left(-1+2\gamma+2\ln\omega+2\ln\left(\frac{b}{2v}\right)\right)}{2v
\sqrt{(eg)^2+(mvb)^2}}
\right]
\nonumber
\eea
where Harmonicnumber$(z)=\Psi(z+1)+\gamma$ and $\gamma$ is the Euler's constant and $\Psi(z)$ is the logarithmic derivative
of $\Gamma$-function.

\section*{Appendix B
\label{appendixb}}
In this appendix we briefly recall some aspects of the PNJL model \cite{PNJL}, which we will then
use to calculate the density of quarks in the medium. This model successfully
describes QCD thermodynamics in the temperature regime we are interested in, by
coupling quarks to the chiral condensate and to a temporal background gauge
field related to the Polyakov loop.
The Euclidean action of the three-flavor PNJL model is
\beq 
{\cal S}_E(\psi, \psi^\dagger, \phi)= \int _0^{\beta=1/T} d\tau\int_V d^3x \left[\psi^\dagger\,\partial_\tau\,\psi + {\cal H}(\psi, \psi^\dagger, \phi)\right] 
-\frac{V}{T}\,\mathcal{U}(\phi,T).
\label{action}
\eeq 
Here $\mathcal{H}$ is the fermionic Hamiltonian density given
by:
\beq
{\cal H} = -i\psi^\dagger\,(\vec{\alpha}\cdot \vec{\nabla}+\gamma_4\,m_0 -\phi)\,\psi + {\cal V}(\psi, \psi^\dagger)~,
\label{H}
\eeq
where $\psi$ is the $N_f=3$ quark field, $\vec{\alpha} = \gamma_0\,
\vec{\gamma}$ and $\gamma_4 = i\gamma_0$ in terms of the standard 
Dirac $\gamma$ matrices and $m_0 = diag(m_{0u},m_{0d},m_{0s})$ is the current quark mass matrix. 
$\mathcal{V}(\psi, \psi^\dagger)$ contains two parts: a
four-fermion interaction acting in the pseudoscalar-isovector/scalar-isoscalar 
quark-antiquark channel, and a six-fermion interaction which
breaks $U_A(1)$ symmetry explicitly:
\bea
\mathcal{V}\left(\psi,\psi^\dagger\right)= &-&\frac
{G}{2}\sum_{f=u,d,s}\left[\left(\bar{\psi}_f\psi_f\right)^2+\left(\bar{\psi}_f\,i\gamma_5
\vec{\tau}\,\psi_f
\right)^2\right]
\nonumber\\
&+&\frac K2\left[\det_{i,j}\left(\bar{\psi}_i\left(1+\gamma_5\right)
\psi_j\right)+\det_{i,j}\left(\bar{\psi}_i\left(1-\gamma_5\right)
\psi_j\right)\right]
\label{V}
\eea

Quarks move in a background color gauge field $\phi \equiv A_4 = iA_0$, 
where $A_0 = \delta_{\mu 0}\,g{\cal A}^\mu_a\,t^a$ with the $SU(3)_c$ gauge 
fields ${\cal A}^\mu_a$ and the generators $t^a = \lambda^a/2$. The matrix 
valued, constant field $\phi$ relates to the (traced) Polyakov loop as follows:
\beq
\Phi=\frac{1}{N_c}\mathrm{Tr}\left[\mathcal{P}\exp\left(i\int_{0}^{\beta}
d\tau A_4\right)\right]=\frac{1}{ 3}\mathrm{Tr}\,e^{i\phi/T}.
\eeq
The thermodynamic potential of the system is:
\bea
&&\!\!\!\!\!\!\!\!\!\!\Omega\left(T,\mu\right)=
\mathcal{U}\left(\Phi,T\right)+\frac{\sigma_{u,d}^{2}}{2G}+\frac{\sigma_{s}^{2}}{4G}-\frac{K}{4G^3}
\sigma_{u,d}^{2}\sigma_s
\\
&&\!\!\!\!\!\!\!\!\!\!-2\sum_{f}\,T\int\frac{\mathrm{d}^3p}{\left(2\pi\right)^3}
\left\{\ln\left[1+\,\Phi e^{-\left(E_{p,f}-\mu_f
\right)/T}+\,\Phi e^{-2\left(E_{p,f}-\mu_f
\right)/T}+\, e^{-3\left(E_{p,f}-\mu_f
\right)/T}\right]\right.
\nonumber\\
&&\!\!\!\!\!\!\!\!\!\!+\left.\ln\left[1+\Phi e^{-\left(E_{p,f}-\bar{\mu}_f
\right)/T}+\Phi e^{-2\left(E_{p,f}-\bar{\mu}_f
\right)/T}+e^{-3\left(E_{p,f}-\bar{\mu}_f
\right)/T}\right]+ 3{E_{p,f}\over T}\theta\left(\Lambda^2-\vec{p}^{~2}\right)\right\}.\nonumber\\
\nonumber
\label{omega2}
\eea
where
\beq
\sigma_i=2G\langle\bar{\psi}_i\psi_i\rangle, ~~~~~~\bar{\mu}_f=-\mu_f,
~~~~~~E_{p,f}=\sqrt{\vec{p}^{~2}+m_{f}^{2}},~~~~~~m_i=m_{0i}-\sigma_i-\frac{K}{4G^2}\sigma_j\sigma_k.
\eeq
By minimizing the thermodynamic potential one can obtain the behavior of the
chiral condensates $\sigma_i$ and of the Polyakov loop $\Phi$ as functions of
the temperature and chemical potential. After this procedure, one is able to
evaluate many thermodynamic quantities.
For example, the quark density we are interested in is the sum of the densities of quarks and
antiquarks and can be obtained through the following formula:
\beq
n_q=\frac{\partial\Omega}{\partial\mu_f}+\frac{\partial\Omega}{\partial\bar{\mu}_f}
\eeq
and it has the following explicit form:
\bea
 n_q\!\!\!\!&\!\!\!=\!\!\!&\!\!\!\!
\frac{N_f}{\pi^2T^3}\!\!\int\!\!\left[\frac{3\exp[\mu/T]\left(\exp[2\mu/T]+\Phi\exp[2E_p/T]+2\Phi
\exp[(E_p+\mu)/T]\right)}{\exp[3E_p/T]+\exp[3\mu/T]+3\Phi\exp[(2E_p+\mu)/T]+3
\Phi\exp[(E_p+2\mu)/T]}\right.
\nonumber\\
&&
\nonumber\\
&&
\nonumber\\
\!\!\!\!\!\!&\!\!\!+&\!\!\!\!\!\!\left.\frac{3\left(1+\Phi\exp[2(E_p+\mu)/T]+2\Phi\exp[(E_p+\mu)/T]\right)}
{1+\exp[3(E_p+\mu)/T]+3\Phi\exp[2(E_p+\mu)/T]+3\Phi\exp[(E_p+\mu)/T]}
\right]p^2dp
\eea
We show the behavior of quark density in Fig. \ref{density}, for typical PNJL model parameters taken from the last 
paper of Ref.~\cite{PNJL}.
\begin{figure}
\begin{center}
\scalebox{.73}{\includegraphics{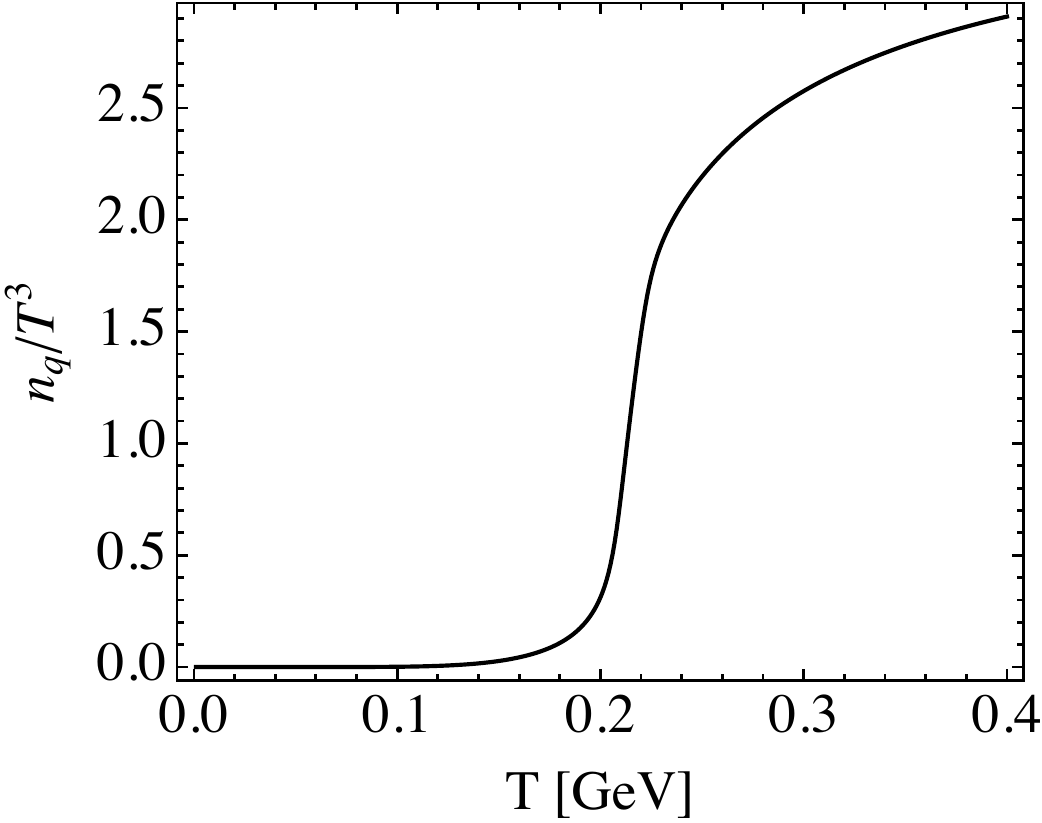}}
\caption{Total quark density as a function of the temperature (PNJL model result).
}
\label{density}
\end{center}
\end{figure}

\end{document}